\documentclass[aps, prd, preprint, nofootinbib, superscriptaddress]{revtex4-2}

\usepackage[utf8]{inputenc}

\usepackage{footnote}
\usepackage{amsmath} \usepackage{amsfonts} \usepackage{amssymb}
\usepackage{wasysym}
\usepackage[dvipsnames]{xcolor}
\usepackage{graphicx}
\usepackage{multirow}
\usepackage{here}
\usepackage{epsfig}
\usepackage{epstopdf}
\usepackage{soul}
\usepackage{hyperref}
\usepackage{graphicx}
\usepackage{verbatim}
\usepackage{booktabs}
\usepackage{xcolor}
\usepackage{array}
\usepackage{dashrule}

\newcommand{\fn}[2]{\mathinner{#1\mathopen{\left(#2\right)}}}

\newcommand{\eq}[1]{Eq.~(\ref{#1})}
\newcommand{\eqs}[2]{Eqs.~(\ref{#1}) and (\ref{#2})}
\newcommand{\eqss}[3]{Eqs.~(\ref{#1}), (\ref{#2}) and (\ref{#3})}

\newcommand{\fref}[1]{FIG.\ref{#1}}
\newcommand{\sref}[1]{Section\ref{#1}}
\newcommand{\tref}[1]{TABLE\ref{#1}}

\newcommand{\TeV}{\mathinner{\mathrm{TeV}}}
\newcommand{\Mpl}{M_\mathrm{pl}}

\begin{document}

\title{Phase Transitions and Gravitational Wave Production at the End of Thermal Inflation}

\author{Hyukjung Kim} 
\affiliation{Lunit Inc., Gangnam-daero, Gangnam-gu 06241, Seoul, Republic of Korea }

\author{İlayda Kuzu}
\affiliation{Department of Physics, Izmir Institute of Technology,
Gülbahçe, Urla 35430, Izmir, Turkiye}

\author{Kerem Özsoy}
\affiliation{Department of Physics, Izmir Institute of Technology,
Gülbahçe, Urla 35430, Izmir, Turkiye}

\author{Zeynep Kahraman}
\affiliation{Department of Physics, Izmir Institute of Technology,
Gülbahçe, Urla 35430, Izmir, Turkiye}

\author{Wan-Il Park} 
\affiliation{Division of Science Education and Institute of Fusion Science, Jeonbuk National University, Jeonju 54896, Korea}

\author{Heeseung Zoe}
\email{heeseungzoe@iyte.edu.tr}
\affiliation{Department of Physics, Izmir Institute of Technology,
Gülbahçe, Urla 35430, Izmir, Turkiye}

\begin{abstract}
We investigate the first-order phase transition that terminates thermal inflation and evaluate the associated stochastic gravitational-wave signals. The transition is first characterized through semi-analytic calculations of the bounce action, which are compared with numerical results obtained using \texttt{CosmoTransitions}. We then study its real-time evolution in a three-dimensional Langevin lattice simulation that incorporates Hubble expansion and the corresponding temperature evolution throughout the transition. The lattice dynamics are consistent with the bounce-action estimates: the transition proceeds through localized bubble nucleation and subsequent bubble growth, rather than through a phase-mixing instability. Using the resulting transition parameters, we estimate the gravitational-wave spectra generated by bubble collisions and acoustic motions in the plasma. The predicted stochastic background lies within the projected sensitivity ranges of future gravitational-wave observatories, including BBO and DECIGO.
\end{abstract}

\maketitle

\newpage

\section{Introduction}

As the early universe cooled, the vacuum structure of particle physics may have evolved through a series of cosmological phase transitions. In the Standard Model (SM), the Higgs potential develops a symmetry-breaking minimum at low temperatures, and the Higgs field settles into this minimum without a strong first-order transition for the observed Higgs mass. In many beyond-the-Standard-Model (BSM) scenarios, however, additional scalar degrees of freedom can generate a potential barrier separating the false and true vacua. The transition then becomes first order and proceeds through the nucleation and subsequent expansion of true-vacuum bubbles.

The nucleation process is induced by quantum or thermal fluctuations around the false vacuum, and its probability is controlled by the corresponding bounce action \cite{Coleman:1977py, Callan:1977pt, Linde:1980tt,Linde:1981zj}. Once nucleated, the bubbles expand, collide, and eventually fill a Hubble volume, completing the transition through cosmological percolation \cite{Guth:1981uk, Kolb:1990vq}. Analytic and semi-analytic treatments usually characterize this process through the bounce action, the nucleation rate, the false-vacuum fraction, and an assumed or estimated bubble wall velocity \cite{Turner:1992tz, Megevand:2016lpr, Ellis:2018mja}. Lattice simulations provide a complementary way to follow the dynamical evolution of the transition beyond these approximations. They make it possible to study bubble nucleation, expansion, collision, and relaxation, and can capture nonlinear transient effects as well as the possible formation of topological defects such as domain walls and cosmic strings \cite{Gleiser:1994cf, Hiramatsu:2012sc, Hindmarsh:2015qta, Cutting:2019zws}.

Cosmological percolation can also generate a stochastic gravitational-wave (GW) background. The released vacuum energy sources gravitational radiation through bubble wall collisions, acoustic waves in the thermal plasma, and magnetohydrodynamic (MHD) turbulence \cite{Kosowsky:1992rz, Caprini:2009yp, Caprini:2015zlo}. The resulting signal offers a potential probe of first-order phase transitions and BSM scalar potentials, motivating searches across a broad frequency range with ground-based interferometers such as Advanced LIGO \cite{LIGOScientific:2014pky}, space-based interferometers such as LISA \cite{LISA:2017pwj} and the proposed Big Bang Observer (BBO) \cite{Harry:2006fi,Corbin:2005ny}, and nanohertz pulsar timing arrays such as NANOGrav \cite{NANOGrav:2023gor} and the International Pulsar Timing Array (IPTA) \cite{Antoniadis:2022pcn}.

Thermal inflation was originally proposed as a solution to the cosmological moduli problem \cite{Lyth:1995hj, Lyth:1995ka}. Many extensions of the SM based on supersymmetric and string theory predict the existence of moduli fields, which can be abundantly produced after primordial inflation and interact only weakly with ordinary matter. The late-time decay of light moduli may therefore spoil the successful predictions of Big Bang nucleosynthesis (BBN). A short period of thermal inflation can dilute the moduli abundance to an acceptable level, while also modifying the pre-existing primordial density fluctuations \cite{Hong:2015oqa,Cho:2017zkj,Hong:2017knn,Bae:2022gkv}. Thermal inflation can also be embedded in scenarios related to Affleck-Dine baryogenesis \cite{Felder:2007iz} and axion production \cite{Kim:2008yu}.

In \cite{Easther:2008sx}, the phase transition and GW production associated with thermal inflation were first investigated. Thermal inflation was shown to have two competing effects on stochastic GWs: it dilutes any pre-existing primordial GW background on small scales, while the phase transition at the end of thermal inflation can generate a new stochastic signal through bubble collisions. By constructing the finite-temperature effective potential and estimating the properties of the bubbles, it was found that the resulting GW spectrum could fall in the frequency range relevant for BBO-like experiments.

The bubble collision picture was later challenged in \cite{Hiramatsu:2014uta}, where thermal fluctuations of the flaton field were incorporated using lattice simulations. It was found that the transition may not proceed through the nucleation, expansion, and percolation of well-defined bubbles. Instead, large thermal fluctuations can drive the field toward the true vacuum through phase mixing, suppressing bubble formation and leading to no appreciable GW signal. However, the lattice simulations were performed at fixed temperature, with the temperature dependence incorporated through a sequence of separate equilibrium simulations rather than through a continuous cosmological evolution.

The issue was revisited more recently in \cite{Dutka:2025oqt}, where bubble nucleation was argued to remain viable in strongly supercooled thermal-inflation-like transitions. In particular, the scalar-field dynamics were studied near the spinodal temperature, where the potential barrier disappears, and it was shown that the field can remain trapped near the origin long enough for critical bubbles to form and expand. This result revives the possibility of a bubble-generated GW signal from thermal inflation. Nevertheless, the lattice dynamics in this approach were not evolved together with the full cosmological background, including continuous Hubble expansion, the temperature dependence of the effective potential, and the subsequent flaton matter-dominated era. This motivates a lattice framework in which the scalar-field dynamics are evolved consistently with the expanding background and the time-dependent thermal potential. The resulting transition history can then be used as input for a separate calculation of the GW spectrum, including the cosmological redshifting induced by the full thermal-inflation history.

In this paper, we perform a comprehensive analysis of the phase transition at the end of thermal inflation and the resulting GW signal. First, we carry out a semi-analytic calculation of the bounce action with an ansatz of the Gaussian profile of the bounce solution and determine the nucleation temperature, evaluating the finite-temperature functions in the effective potential by Gaussian quadrature. Second, we perform lattice simulations in which the Langevin dynamics of the flaton field are evolved together with Hubble expansion and the time-dependent thermal potential. Finally, we compute the GW spectrum from the resulting transition history, taking into account the additional redshifting associated with the post-transition flaton matter-dominated era.

This paper is organized as follows. In \sref{sec:fdti}, we present the basic equations governing the background dynamics of thermal inflation. In \sref{sec:pteti}, we describe the phase transition at the end of thermal inflation and compute the bounce action using a semi-analytic approach. In \sref{sec:ns}, we present numerical simulations of the tunneling dynamics and lattice evolution, and examine the consistency between the semi-analytic and lattice descriptions. In \sref{sec:gwb}, we calculate the resulting GW spectrum and discuss its observability. Finally, in \sref{sec:conclusion}, we summarize our results and present our conclusions.

\section{Flaton dynamics in thermal inflation}\label{sec:fdti}

Thermal inflation is realized when a scalar field, called the flaton, is temporarily trapped near the origin by finite-temperature effects in an otherwise nearly flat potential. The false-vacuum energy then drives a short period of accelerated expansion. As the universe cools, the thermal stabilization becomes weaker, and the flaton eventually leaves the origin, ending thermal inflation. In this section, we formulate the flaton dynamics, identify the characteristic temperatures of the thermal history, and introduce the stochastic real-time dynamics used in the lattice simulations.

\subsection{Finite-temperature effective potential of the flaton}

To describe the flaton dynamics in thermal inflation, we use the finite-temperature effective potential parameterized as in \cite{Easther:2008sx,Dolan:1973qd},
\begin{equation}
\begin{split}
\fn{V}{\phi,T}
&=
\fn{V_\mathrm{vac}}{\phi}
+
\fn{V_\mathrm{th}}{\phi,T}
\\
&=
V_0
-\frac{1}{2}m^2\phi^2
+\frac{1}{4}\lambda\phi^4
+
T^4
\left[
n_f\fn{J_F}{\dfrac{m_F^2}{T^2}}
+
n_b\fn{J_B}{\dfrac{m_B^2}{T^2}}
\right].
\end{split}
\label{eq:fullV}
\end{equation}
Here $n_f$ and $n_b$ denote the effective numbers of fermionic and bosonic degrees of freedom coupled to the flaton. The thermal functions are defined by
\begin{align}
\fn{J_F}{u^2}
&=
-\frac{1}{2\pi^2}
\int_0^\infty dx\,x^2
\log\left(
1+e^{-\sqrt{u^2+x^2}}
\right),
\\
\fn{J_B}{u^2}
&=
+\frac{1}{2\pi^2}
\int_0^\infty dx\,x^2
\log\left(
1-e^{-\sqrt{u^2+x^2}}
\right).
\label{eq:Jfdef}
\end{align}
The field-dependent thermal masses are taken to be \cite{ Easther:2008sx, Hiramatsu:2014uta}
\begin{align}
m_B^2(\phi,T)
&=
m_b^2
+
\frac{1}{2}y_b^2\phi^2
+
\left(
\frac{1}{4}y_b^2
+
\frac{2}{3}g^2
\right)T^2,
\\
m_F^2(\phi,T)
&=
\frac{1}{2}y_f^2\phi^2
+
\frac{1}{6}g^2T^2,
\label{eq:meff}
\end{align}
where $y_b$, $y_f$, and $g$ are dimensionless couplings.  Assuming the scalar mass is positive at the high-energy input scale, these couplings are typically chosen to be of order unity to generate the substantial radiative corrections necessary to drive the flaton mass-squared negative. The symbols $B$ and $F$ refer to the bosonic and fermionic thermal contributions, respectively. At the origin, it is useful to define the dimensionless arguments of the thermal functions as
\begin{equation}
\delta_f
\equiv
\frac{m_F^2(0,T)}{T^2}
=
\frac{1}{6}g^2,
\qquad
\fn{\delta_b}{T}
\equiv
\frac{m_B^2(0,T)}{T^2}
=
\frac{m_b^2}{T^2}
+
\frac{1}{4}y_b^2
+
\frac{2}{3}g^2 .
\label{eq:delta_def}
\end{equation}

At zero temperature, the flaton potential has a minimum at a large field value, which we denote by $\phi_0$. We parametrize this vacuum expectation value as
\begin{equation}
\gamma
\equiv
\frac{\phi_0}{\Mpl}.
\end{equation}
For the tree-level potential in \eq{eq:fullV}, the condition that the minimum is located at $\phi=\phi_0$ gives $\lambda=m^2/\phi_0^2$. Since the typical vacuum expectation value for thermal inflation lies in the intermediate scale range $\phi_0 \sim 10^7 \text{--} 10^{9} \text{ TeV}$, the corresponding quartic coupling $\lambda$ is extremely small, rendering the zero-temperature potential exceptionally flat. Requiring the vacuum energy to vanish at the true minimum then fixes $V_0=m^2\phi_0^2/4$. Therefore,
\begin{equation}
\lambda
=
\frac{m^2}{\phi_0^2}
=
\frac{m^2}{\gamma^2\Mpl^2},
\qquad
V_0
=
\frac{1}{4}m^2\phi_0^2
=
\frac{1}{4}\gamma^2m^2\Mpl^2 .
\label{eq:pot_params}
\end{equation}
With this zero-temperature structure, finite-temperature corrections stabilize the flaton near the origin at high temperature, producing a false-vacuum configuration whose vacuum energy drives thermal inflation. As the temperature decreases, the effective potential evolves and eventually allows the flaton to leave the origin and move toward the true vacuum.

\begin{figure}[!ht]
    \centering
    \includegraphics[width=1\linewidth]{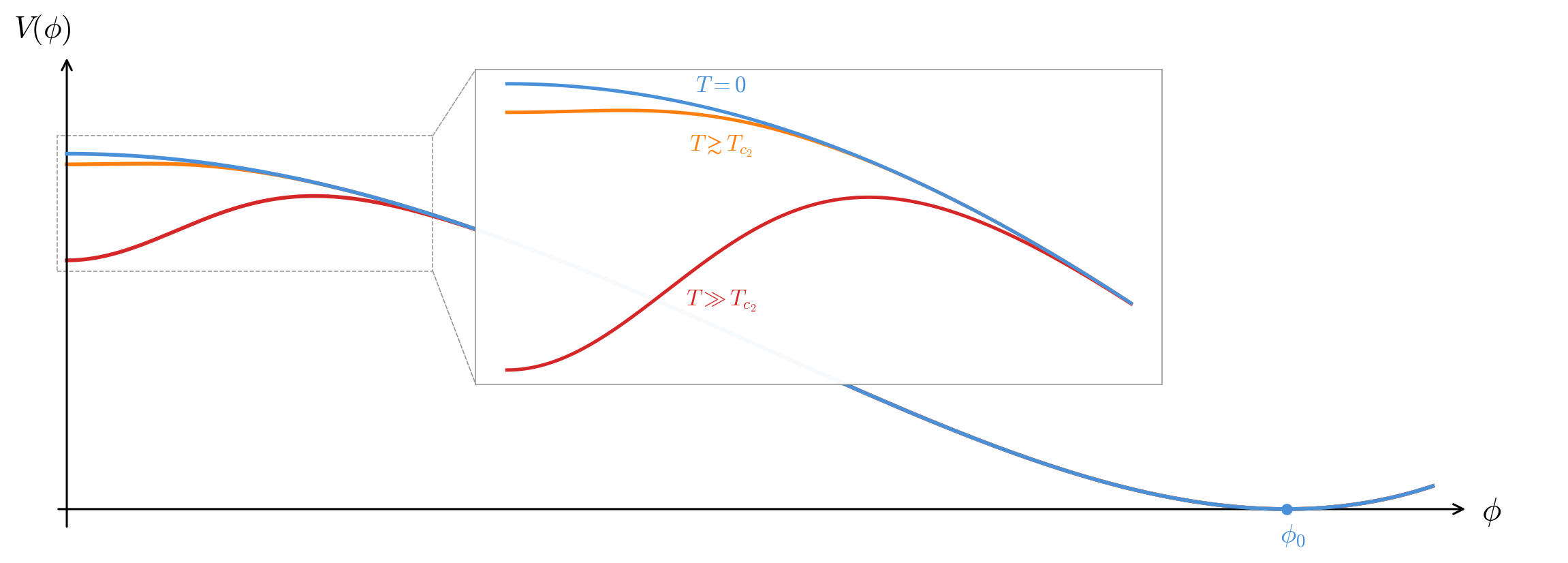}
    \caption{A schematic picture of the finite-temperature effective potential of the flaton field. Finite-temperature corrections hold the flaton near the origin, where the false-vacuum energy $V_0$ dominates and drives thermal inflation. As the temperature decreases, the barrier separating the false and true vacua becomes shallower, eventually allowing the transition to the true vacuum.}
    \label{fig:flaton_thermal_potential}
\end{figure}

\subsection{Beginning and end of thermal inflation}

We now describe the thermal history of thermal inflation in terms of the characteristic temperatures that mark its main stages. Thermal inflation begins at the temperature where the false-vacuum energy starts to dominate the total energy density. The accelerated expansion during this stage dilutes the pre-existing moduli abundance. As the universe cools further, the potential barrier between the false and true vacua becomes weaker, and a first-order phase transition terminates thermal inflation at $T_\mathrm{c_1}$.

To estimate the onset temperature more explicitly, we compare the relevant energy components before thermal inflation begins. We denote by $a_\mathrm{a}$ the scale factor at the epoch when the moduli matter and radiation energy densities are comparable, and by $\rho_0$ their common value at that time. While the flaton is trapped near the origin, the total energy density is
\begin{equation}
\rho(a,T)
=
\rho_\mathrm{m}
+
\rho_\mathrm{r}
+
\rho_\phi
=
\rho_0
\left(
\frac{a_\mathrm{a}}{a}
\right)^3
+
\rho_0
\left(
\frac{a_\mathrm{a}}{a}
\right)^4
+
\fn{V}{0,T}.
\label{eq:rho_components_ti}
\end{equation}
Before thermal inflation, the universe is assumed to be moduli dominated. Therefore, the ratio of matter to radiation at the beginning of thermal inflation is estimated as
\begin{equation}
\left.
\frac{\rho_\mathrm{m}}{\rho_\mathrm{r}}
\right|_{t=t_\mathrm{b}}
\sim
\frac{a_\mathrm{b}}{a_\mathrm{a}}
\sim
\left(
\frac{H_\mathrm{a}}{H_\mathrm{b}}
\right)^{2/3}
\sim
\left[
\frac{m_s^2\Mpl^2}{\fn{V}{0,T_\mathrm{b}}}
\right]^{1/3} \sim \gamma^{-2/3} \gg 1.
\label{eq:ratio1}
\end{equation}
On the other hand, thermal inflation begins when the false-vacuum energy becomes comparable to the moduli matter density. The same ratio can be written as
\begin{equation}
\left.
\frac{\rho_\mathrm{m}}{\rho_\mathrm{r}}
\right|_{t=t_\mathrm{b}}
\sim
\frac{\fn{V}{0,T_\mathrm{b}}}
{\dfrac{\pi^2}{30}\fn{g_*}{T_\mathrm{b}}T_\mathrm{b}^4} \gg 1.
\label{eq:ratio2}
\end{equation}
Combining \eqss{eq:pot_params}{eq:ratio1}{eq:ratio2}, we obtain
\begin{equation}
T_\mathrm{b}
\simeq
\left[
\frac{30}{\pi^2\fn{g_*}{T_\mathrm{b}}}
\right]^{1/4}
\frac{\fn{V}{0,T_\mathrm{b}}^{1/3}}
{m_s^{1/6}\Mpl^{1/6}} .
\label{eq:Tb_general}
\end{equation}
Using
\begin{equation}
\fn{V}{0,T_\mathrm{b}}
=
V_0
+
T_\mathrm{b}^4
\left[
n_f\fn{J_F}{\delta_f}
+
n_b\fn{J_B}{\fn{\delta_b}{T_\mathrm{b}}}
\right],
\label{eq:V_origin_Tb}
\end{equation}
and expanding in the thermal correction to $V_0$, this gives
\begin{align}
\fn{T_\mathrm{b}}{\gamma}
&\simeq
\left[
\frac{30}{\pi^2\fn{g_*}{T_\mathrm{b}}}
\right]^{1/4}
\frac{V_0^{1/3}}{m_s^{1/6}\Mpl^{1/6}}
\left[
1
+
\frac{
T_\mathrm{b}^4
\left(
n_f\fn{J_F}{\delta_f}
+
n_b\fn{J_B}{\fn{\delta_b}{T_\mathrm{b}}}
\right)
}{3V_0}
\right]
\nonumber
\\
&=
4^{-1/3}
\left[
\frac{30}{\pi^2\fn{g_*}{T_\mathrm{b}}}
\right]^{1/4}
\gamma^{2/3}
\left(
\frac{m}{m_s}
\right)^{1/6}
\sqrt{m\Mpl}
\left[
1
+
\frac{
T_\mathrm{b}^4
\left(
n_f\fn{J_F}{\delta_f}
+
n_b\fn{J_B}{\fn{\delta_b}{T_\mathrm{b}}}
\right)
}{3V_0}
\right].
\label{eq:Tb_gamma}
\end{align}

During thermal inflation, $T_\mathrm{c_1}\lesssim T\lesssim T_\mathrm{b}$, the flaton remains trapped near the origin. Assuming that the temperature scales as $T\propto a^{-1}$ during this period, the Hubble rate can be written as
\begin{equation}
\fn{H^2}{T}
\simeq
\frac{1}{3\Mpl^2}
\left[
\fn{V}{0,T_\mathrm{b}}
\left(
\frac{T}{T_\mathrm{b}}
\right)^3
+
\rho_\mathrm{r}(T_\mathrm{b})
\left(
\frac{T}{T_\mathrm{b}}
\right)^4
+
\fn{V}{0,T}
\right],
\label{eq:H_TI_compact}
\end{equation}
where
\begin{equation}
\rho_\mathrm{r}(T_\mathrm{b})
=
\frac{\pi^2}{30}\fn{g_*}{T_\mathrm{b}}T_\mathrm{b}^4 .
\label{eq:rho_r_Tb}
\end{equation}
Equivalently, using \eq{eq:Tb_general}, the radiation term at $T_\mathrm{b}$ can be estimated as
\begin{equation}
\rho_\mathrm{r}(T_\mathrm{b})
\sim
\frac{\fn{V}{0,T_\mathrm{b}}^{4/3}}
{m_s^{2/3}\Mpl^{2/3}} .
\label{eq:rho_r_Tb_estimate}
\end{equation}

Thermal inflation begins at $T=T_\mathrm{b}$ and, as discussed in detail in the following sections, ends through a first-order phase transition at $T=T_\mathrm{c_1}$. Independently of this tunneling process, the finite-temperature potential defines another characteristic temperature, denoted by $T_\mathrm{c_2}$, at which the curvature at $\phi=0$ changes sign. In the absence of bubble nucleation, $T_\mathrm{c_2}$ would correspond to the spinodal temperature, where the origin becomes unstable and the field rolls toward the true vacuum. In the parameter region considered below, we find $T_\mathrm{c_1}\sim T_\mathrm{c_2}$. We now determine $T_\mathrm{c_2}$ from the curvature of the finite-temperature effective potential.

Expanding the potential around $\phi=0$, we obtain
\begin{align}
\fn{V}{\phi,T}
&\simeq
V_0
+
T^4
\left[
n_f\fn{J_F}{\delta_f}
+
n_b\fn{J_B}{\fn{\delta_b}{T}}
\right]
\nonumber
\\
&\quad
+
\frac{1}{2}
\left[
-m^2
+
n_f y_f^2 \fn{J_F^\prime}{\delta_f}T^2
+
n_b y_b^2 \fn{J_B^\prime}{\fn{\delta_b}{T}}T^2
\right]\phi^2
+
\cdots .
\label{eq:potential_expansion_origin}
\end{align}
The spinodal temperature $T_\mathrm{c_2}$ is determined by the vanishing of the effective mass squared at the origin,
\begin{equation}
T_\mathrm{c_2}
=
\frac{m}
{
\sqrt{
n_f y_f^2 \fn{J_F^\prime}{\delta_f}
+
n_b y_b^2 \fn{J_B^\prime}{\fn{\delta_b}{T_\mathrm{c_2}}}
}
}.
\label{eq:Tc2}
\end{equation}
Because $\fn{\delta_b}{T}$ depends on temperature through the bosonic mass term, \eq{eq:Tc2} must in general be solved numerically. In the fermion-only limit, it reduces to
\begin{equation}
T_\mathrm{c_2}
=
\frac{m}
{\sqrt{n_f y_f^2 \fn{J_F^\prime}{\delta_f}}}.
\label{eq:Tc2_fer}
\end{equation}

Neglecting changes in $g_*$ during thermal inflation, the number of e-folds is estimated as
\begin{equation}
N_\mathrm{TI}
=
\log\left(
\frac{a_\mathrm{c_1}}{a_\mathrm{b}}
\right)
\simeq
\log\left(
\frac{T_\mathrm{b}}{T_\mathrm{c_1}}
\right)
\simeq
\log\left(
\frac{T_\mathrm{b}}{T_\mathrm{c_2}}
\right),
\label{eq:N_TI}
\end{equation}
where the last approximation uses $T_\mathrm{c_1}\sim T_\mathrm{c_2}$.

\subsection{Thermal fluctuations and flaton dynamics}\label{subsec:flaton_dynamics}

The flaton evolves in a thermal environment whose temperature changes continuously as the universe expands. Interactions with the thermal bath generate stochastic fluctuations of the flaton field, while the same interactions also induce dissipation. To describe this real-time dynamics, we use a Langevin-type equation for $\phi$,
\begin{align}
\ddot{\phi}
+
(\eta+3H)\dot{\phi}
=
\frac{1}{a^2}\nabla^2\phi
-
\frac{\partial V(\phi,T)}{\partial\phi}
+
\xi(\mathbf{x},t),
\label{eq:langevin}
\end{align}
where $\mathbf{x}$ denotes the comoving spatial coordinate, $t$ is cosmic time, and the derivative of the potential is taken with respect to $\phi$ at fixed instantaneous temperature $T(t)$. The stochastic source $\xi$ is taken to be Gaussian thermal noise \cite{Figueroa:2021yhd,Dutka:2025oqt,Rodrigues:2025neh}, satisfying
\begin{align}
\left\langle \xi(\mathbf{x},t) \right\rangle
&=
0,
\\
\left\langle
\xi(\mathbf{x},t)
\xi(\mathbf{x}',t')
\right\rangle
&=
\frac{D(T)}{a^3}
\delta^{(3)}(\mathbf{x}-\mathbf{x}')
\delta(t-t') .
\label{eq:noise_corr}
\end{align}
The fluctuation-dissipation relation gives
\begin{equation}
D(T)=2\eta T ,
\end{equation}
where $\eta$ is the damping coefficient. In the numerical analysis, $\eta$ is taken to be of order the temperature. Because the noise amplitude and the dissipative term are related by the fluctuation-dissipation relation, changing $\eta$ affects the relaxation time scale but does not shift the thermal equilibrium configuration.

Embedding the Langevin dynamics in an expanding background introduces two important effects. First, the spatial-gradient term is suppressed by the factor $1/a^2$. Thus, spatial inhomogeneities are redshifted as the universe expands, and the competition between gradient energy and potential energy is modified during bubble formation and growth. Second, the stochastic noise is diluted both by the decrease of the temperature, $T(t)\propto a^{-1}(t)$, and by the volume factor $1/a^3$ in the noise correlator. As a result, thermal fluctuations become progressively less efficient at driving large field excursions as thermal inflation proceeds. These effects are absent in fixed-temperature lattice simulations and are essential for following the phase transition in a cosmological background.

\section{Phase transitions at the end of thermal inflation}\label{sec:pteti}

Thermal inflation can terminate while the finite-temperature potential barrier still separates the false and true vacua by getting over the potential barrier through thermal fluctuation. In this case, true-vacuum bubbles nucleate before the origin becomes spinodally unstable at $T_\mathrm{c_2}$. Since the background is rapidly expanding during thermal inflation, the relevant question is whether the tunneling rate becomes sufficiently large compared with the Hubble expansion rate. We determine the nucleation temperature $T_\mathrm{n}$ using the semiclassical bounce formalism \cite{Coleman:1977py,Callan:1977pt} and its finite-temperature extension, in which the tunneling exponent is controlled by the three-dimensional Euclidean action $S_3$ \cite{Linde:1980tt,Linde:1981zj}.

In the present model, the finite-temperature effective potential contains the thermal functions $J_B$ and $J_F$, which makes it difficult to obtain $S_3(T)$ and $T_\mathrm{n}$ in closed analytic form. We therefore evaluate the thermal functions using a Gaussian quadrature approximation around the critical temperature in Appendix\ref{sec:gqm} and use the resulting potential to compute the bounce action semi-analytically. This treatment follows the standard finite-temperature conventions for the thermal functions used in phase-transition calculations \cite{Fowlie:2018eiu}. 
The resulting estimate of $T_\mathrm{n}$ provides a semi-analytic benchmark against which the Langevin and lattice simulations of the next section can be compared.

\subsection{Bounce action}
 
To obtain a semi-analytic estimate of the bounce action, we introduce the Gaussian ansatz
\begin{equation}
\phi(r)
=
\phi_\mathrm{e}
\exp\left(
-\frac{r^2}{R^2}
\right),
\label{eq:gaussian_ansatz}
\end{equation}
and define
\begin{equation}
\alpha_f
\equiv
\frac{y_f^2\phi_\mathrm{e}^2}{2T^2},
\qquad
\alpha_b
\equiv
\frac{y_b^2\phi_\mathrm{e}^2}{2T^2}.
\label{eq:alpha_def}
\end{equation}
With this ansatz, the three-dimensional Euclidean action becomes
\begin{equation}
S_3(\phi_\mathrm{e},R)
=
\frac{3\pi^{3/2}}{4\sqrt{2}}
\phi_\mathrm{e}^2R
+
R^3U(\phi_\mathrm{e},T),
\label{eq:S3_gaussian}
\end{equation}
with
\begin{equation}
U(\phi_\mathrm{e},T)
=
-\frac{\pi^{3/2}}{4\sqrt{2}}m^2\phi_\mathrm{e}^2
+
\frac{\pi^{3/2}}{32}\lambda\phi_\mathrm{e}^4
+
n_fT^4F_F(\alpha_f,\delta_f)
+
n_bT^4F_B(\alpha_b,\delta_b(T)).
\label{eq:U_gaussian}
\end{equation}
\begin{equation}
F_F(\alpha_f,\delta_f)
\equiv
2\pi
\int_0^\infty dx\,x^{1/2}
\left[
J_F\left(\delta_f+\alpha_f e^{-2x}\right)
-
J_F(\delta_f)
\right]
\end{equation}
\begin{equation}
F_B(\alpha_b,\delta_b)
\equiv
2\pi
\int_0^\infty dx\,x^{1/2}
\left[
J_B\left(\delta_b+\alpha_b e^{-2x}\right)
-
J_B(\delta_b)
\right]
\end{equation}

The high- and low-temperature expansions of the thermal functions $J_F$ and $J_B$ are not reliable near the critical temperature relevant for the first-order phase transition. We therefore evaluate these functions using the Gaussian quadrature method summarized in Appendix\ref{sec:gqm}.
The first-order phase transition occurs around $T\sim m$, as supported by the numerical simulations. Around this temperature, the quadrature approximation gives
\begin{equation}
F_X(\alpha,\delta)
\simeq
\frac{\pi}{\sqrt{2}}
\sum_{i=1}^{2}
w_i
\frac{
J_X(\delta+x_i\alpha)-J_X(\delta)
}{x_i},
\qquad
X=F,B ,
\label{eq:F_quadrature_main}
\end{equation}
with
\begin{align}
x_1&=0.1535900739,
&
w_1&=0.5563908709,
\nonumber
\\
x_2&=0.6908657079,
&
w_2&=0.3298360546 .
\label{eq:wixi_main}
\end{align}
The corresponding derivative is
\begin{equation}
F_X^\prime(\alpha,\delta)
\simeq
\frac{\pi}{\sqrt{2}}
\sum_{i=1}^{2}
w_i
J_X^\prime(\delta+x_i\alpha),
\qquad
X=F,B ,
\label{eq:Fprime_quadrature_main}
\end{equation}
where the prime denotes differentiation with respect to $\alpha$.

The stationary configuration is obtained by varying the action with respect to $R$ and $\phi_\mathrm{e}$:
\begin{align}
\frac{3\pi^{3/2}}{4\sqrt{2}}\phi_\mathrm{e}^2
+
3R^2U
&=
0,
\\
\frac{3\pi^{3/2}}{2\sqrt{2}}\phi_\mathrm{e}
+
R^2U_{,\phi_\mathrm{e}}
&=
0.
\label{eq:variation_gaussian}
\end{align}
Since $\lambda=m^2/\phi_0^2$ is extremely small in thermal-inflation models, we neglect the quartic contribution in the variational equations. Then
\begin{equation}
\phi_\mathrm{e}^2
=
\frac{2\alpha_fT^2}{y_f^2},
\label{eq:phi_e_alpha}
\end{equation}
where $\alpha_f$ is determined by
\begin{align}
\frac{m^{2}}{y_{f}^{2}T^{2}}
&=
\frac{n_f}{\sqrt{\pi}}
\sum_{i=1}^{2}w_i
\left[
\frac{3}{\alpha_f}
\frac{
J_F(\delta_f+x_i\alpha_f)-J_F(\delta_f)
}{x_i}
-
J_F^\prime(\delta_f+x_i\alpha_f)
\right]
\nonumber
\\
&\quad
+
\frac{n_b}{\sqrt{\pi}}
\sum_{i=1}^{2}w_i
\left[
\frac{3}{\alpha_f}
\frac{
J_B\left(\delta_b(T)+x_i\frac{y_b^2}{y_f^2}\alpha_f\right)
-
J_B\left(\delta_b(T)\right)
}{x_i}
-
\frac{y_b^2}{y_f^2}
J_B^\prime
\left(
\delta_b(T)+x_i\frac{y_b^2}{y_f^2}\alpha_f
\right)
\right].
\label{eq:alpha_f_equation}
\end{align}
We also define $\alpha_b$ as
\begin{equation}
\alpha_b
=
\frac{y_b^2}{y_f^2}\alpha_f .
\label{eq:alpha_b_alpha_f}
\end{equation}

The critical radius is
\begin{equation}
R^{2}
=
-\frac{\pi^{3/2}}{\sqrt{2}}
\frac{1}
{T^{2}
\left\{
n_fy_f^2
\left[
F_F^\prime(\alpha_f,\delta_f)
-
\frac{F_F(\alpha_f,\delta_f)}{\alpha_f}
\right]
+
n_by_b^2
\left[
F_B^\prime(\alpha_b,\delta_b(T))
-
\frac{F_B(\alpha_b,\delta_b(T))}{\alpha_b}
\right]
\right\}
}.
\label{eq:R2_semi_analytic}
\end{equation}
The bounce action is then given by
\begin{equation}
\frac{S_3}{T}
=
\frac{\pi^{3/2}}{\sqrt{2}}
\frac{\alpha_f}{y_f^2}
\sqrt{
-\frac{\pi^{3/2}}{\sqrt{2}}
\frac{1}
{
n_fy_f^2
\left[
F_F^\prime(\alpha_f,\delta_f)
-
\frac{F_F(\alpha_f,\delta_f)}{\alpha_f}
\right]
+
n_by_b^2
\left[
F_B^\prime(\alpha_b,\delta_b(T))
-
\frac{F_B(\alpha_b,\delta_b(T))}{\alpha_b}
\right]
}
}.
\label{eq:S3_T_semi_analytic}
\end{equation}
Equations \eq{eq:alpha_f_equation} and \eq{eq:S3_T_semi_analytic} provide the semi-analytic estimate of $S_3/T$ used to determine the nucleation temperature.

\subsection{Nucleation temperature}

The nucleation temperature $T_\mathrm{n}$ is determined by the condition that the bubble nucleation rate becomes comparable to the Hubble expansion rate. Using the standard estimate \cite{Linde:1980tt}
\begin{equation}
\fn{\Gamma}{T}=
\frac{T^4}{H^4(T)}
\left(
\frac{S_3(T)}{2\pi T}
\right)^{3/2}
\exp\left[
-\frac{S_3(T)}{T}
\right]. \label{eq:gamma}
\end{equation}
Note that the standard estimate for the nucleation prefactor in \eqref{eq:gamma} carries theoretical uncertainty near the spinodal temperature. While an exact calculation could yield a more precise rate, its effect is strictly subdominant compared to the exponential contribution; therefore, we retain the standard approximation.

We define $T_\mathrm{n}$ by
\begin{equation}
\frac{\fn{\Gamma}{T_\mathrm{n}}}{H^4}
=
\frac{T_\mathrm{n}^4}{H^4}\left(
\frac{S_3(T)}{2\pi T}
\right)^{3/2}
\exp\left[
-\frac{\fn{S_3}{T_\mathrm{n}}}{T_\mathrm{n}}
\right]
\sim
1 .
\label{eq:Tndef}
\end{equation}
During thermal inflation, the background energy density is dominated by the false-vacuum energy. Thus, using \eq{eq:pot_params}, the Hubble parameter is approximately
\begin{equation}
H_\mathrm{TI}
=
\sqrt{\frac{V_0}{3\Mpl^2}}
=
\frac{\gamma m}{2\sqrt{3}} .
\label{eq:Hubble}
\end{equation}
From \eqs{eq:Tndef}{eq:Hubble}, we define the dimensionless function
\begin{equation}
\fn{C}{T}
\equiv
\frac{\fn{S_3}{T}}{T}
+\dfrac{3}{2}\ln\left(\dfrac{S_3(T)}{2\pi T}\right)-
4\fn{\ln}{\frac{2\sqrt{3}T}{\gamma m}} .
\label{eq:C_T}
\end{equation}
The nucleation condition is then
\begin{equation}
\fn{C}{T_\mathrm{n}}
=
0 .
\label{eq:nuclcond}
\end{equation}

\begin{table}[h!]
\centering
\begin{tabular}{cccccc}
\hline
$T \ (\TeV)$ & $\alpha_f(T)$ & $\alpha_b(T)$ & $S_3/T$ & $4 \ln(2\sqrt{3}T/\gamma m)$ & $C(T)$ \\  \hline
1.14 & --      & --      & --        & 73.46815 & --        \\
1.15 & --      & --      & --        & 73.50309 & --        \\
1.16 & 0.41567 & 0.41567 & 13.89133  & 73.53772 & -58.45631 \\
1.17 & 0.94999 & 0.94999 & 22.57239  & 73.57205 & -49.08139 \\
1.18 & 1.50408 & 1.50408 & 30.12550  & 73.60610 & -41.12936 \\
1.19 & 2.07716 & 2.07716 & 37.27157  & 73.63985 & -33.69775 \\
1.20 & 2.66840 & 2.66840 & 44.25269  & 73.67333 & -26.49257 \\
1.21 & 3.27710 & 3.27710 & 51.18342  & 73.70652 & -19.37679 \\
1.22 & 3.90268 & 3.90268 & 58.12793  & 73.73944 & -12.27436 \\
1.23 & 4.54462 & 4.54462 & 65.12608  & 73.77210 & -5.13835  \\
1.24 & 5.20248 & 5.20248 & 72.20446  & 73.80448 & 2.06241  \\
1.25 & 5.87585 & 5.87585 & 79.38176  & 73.83661 & 9.34974   \\
1.26 & 6.56435 & 6.56435 & 86.67166  & 73.86849 & 16.73955  \\
1.27 & 7.26764 & 7.26764 & 94.08450  & 73.90011 & 24.24386  \\
1.28 & 7.98540 & 7.98540 & 101.62830 & 73.93148 & 31.87199 
\end{tabular}
\caption{Temperature dependence of $\alpha_f(T)$, $\alpha_b(T)$, $S_3/T$, and the nucleation criterion $C(T)$ near the end of thermal inflation. The nucleation temperature is determined by $C(T_\mathrm{n})=0$.}
\label{table:nucleation}
\end{table}

\tref{table:nucleation} shows the result of the semi-analytic calculation based on \eqs{eq:S3_T_semi_analytic}{eq:C_T} for the benchmark parameter set
\begin{equation}
\gamma=4.1667\times10^{-8},
\qquad
n_b=n_f=20,
\qquad
y_b=y_f=1.09,
\qquad
g=1.05 .
\end{equation}
$T_n$ is found to be
\begin{equation}
T_\mathrm{n}
\simeq
1.237\TeV .
\end{equation}

\begin{figure}[!h]
    \centering
    \includegraphics[width=0.7\linewidth]{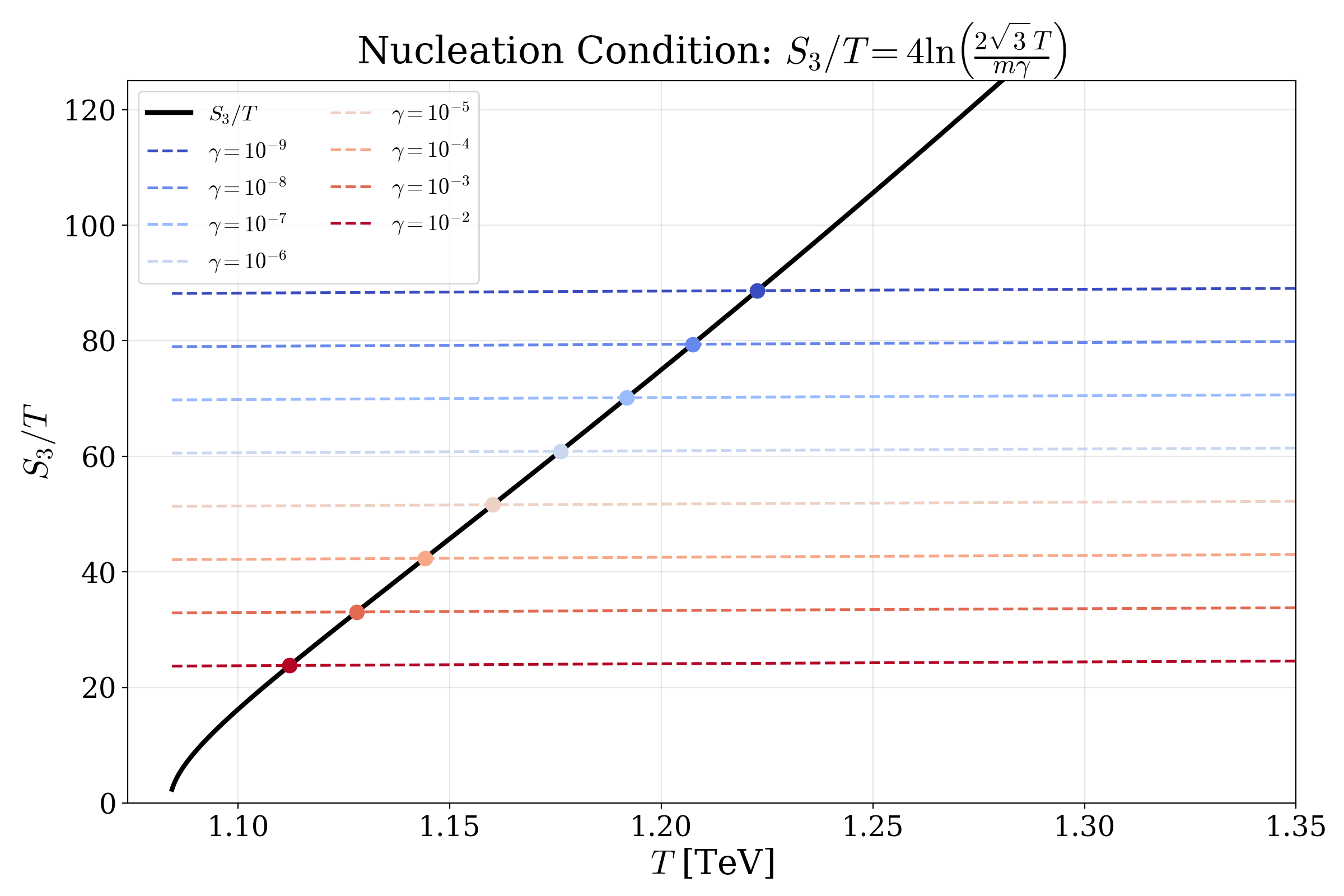}
    \caption{Dependence of the nucleation temperature $T_\mathrm{n}$ on $\gamma$. The other parameters are fixed at $n_b=20$, $n_f=20$, $y_b=y_f=1.09$, and $g=1.05$.}
    \label{fig:T_n_with_gamma}
\end{figure}

It is useful to make the dependence of $T_\mathrm{n}$ on $\gamma$ explicit. Defining
\begin{equation}
\fn{\mathcal{S}}{T_\mathrm{n}}
\equiv
\frac{\fn{S_3}{T_\mathrm{n}}}{T_\mathrm{n}},
\end{equation}
since $\mathcal{S}(T_n)\gg1$ near $T_n$, $\fn{\mathcal{S}}{T_\mathrm{n}}-\dfrac{3}{2}\ln\left(\dfrac{\mathcal{S}(T_n)}{2\pi}\right)\simeq\mathcal{S}(T_n)$. Thus, the nucleation condition can be rewritten as
\begin{equation}
\fn{\mathcal{S}}{T_\mathrm{n}}
=
4\fn{\ln}{\frac{2\sqrt{3}T_\mathrm{n}}{m}}
-
4\fn{\ln}{\gamma}.
\label{eq:nucl_gamma_exact}
\end{equation}
Since $T_\mathrm{n}$ appears in the first term only logarithmically, the dominant dependence on $\gamma$ is controlled by the term $-4\fn{\ln}{\gamma}$. We therefore fit the numerical solution of the exact condition \eq{eq:nuclcond} by
\begin{equation}
T_\mathrm{n}[\mathrm{GeV}]
\simeq
A_0
+
A_1\fn{\ln}{\gamma}
+
A_2\left(\fn{\ln}{\gamma}\right)^2 .
\label{eq:Tn_func}
\end{equation}
For the parameter range shown in \fref{fig:T_n_with_gamma}, the best-fit coefficients are
\begin{equation}
A_0=1077.02,
\qquad
A_1=-7.4794,
\qquad
A_2=-0.022111 .
\end{equation}
This parametric dependence of $T_\mathrm{n}$ on $\gamma$ is illustrated in \fref{fig:T_n_with_gamma}.

\section{Numerical simulations}\label{sec:ns}

\subsection{Bounce-action calculation}

In this section, we numerically analyze the tunneling and percolation dynamics at the end of thermal inflation. We compute the bounce action $S_3(T)/T$ using the \texttt{CosmoTransitions} package \cite{Wainwright:2011kj}. Using the resulting temperature-dependent action, we evaluate the nucleation rate and the quantities that characterize the progress of the phase transition \cite{Wang:2020jrd}:
\begin{align}
\frac{\fn{\Gamma}{T}}{H^4(T)}
&=
\frac{T^4}{H^4(T)}
\left(
\frac{S_3(T)}{2\pi T}
\right)^{3/2}
\exp\left[
-\frac{S_3(T)}{T}
\right],
\label{eq:tun_hub}
\\
n(T)
&=
\int_T^{T_\mathrm{c}}
dT'\,
\frac{\Gamma(T')}
{T'H^4(T')},
\label{eq:nuc_temp}
\\
I(T)
&=
\frac{4\pi}{3}
\int_T^{T_\mathrm{c}}
d\bar{T}\,
\frac{\Gamma(\bar{T})}
{\bar{T}^4H(\bar{T})}
\left[
\int_T^{\bar{T}}
dT'\,
\frac{v_\mathrm{w}(T')}{H(T')}
\right]^3,
\nonumber
\\
P(T)
&=
\exp\left[
-I(T)
\right].
\label{eq:percol}
\end{align}
Here $T_\mathrm{c}$ denotes the temperature at which the false and true vacua are degenerate, $n(T)$ is the expected number of nucleated bubbles per Hubble volume, and $P(T)$ is the fraction of space remaining in the false vacuum. These expressions assume adiabatic cooling, $T\propto a^{-1}$, with approximately constant effective numbers of relativistic degrees of freedom. In the numerical analysis below, we adopt the relativistic-wall approximation and set $v_\mathrm{w}=1$.

We extract three characteristic temperatures from these quantities. The nucleation temperature $T_\mathrm{n}$ is defined by
\begin{equation}
n(T_\mathrm{n})
=
1.
\end{equation}
The percolation temperature $T_\mathrm{p}$ is determined by
\begin{equation}
I(T_\mathrm{p})
=
0.34,
\qquad
P(T_\mathrm{p})
=
e^{-0.34}
\simeq
0.71.
\end{equation}
where the fraction of the true vacuum volume exceeds $0.71$. We also introduce a practical completion temperature $T_\mathrm{c_1}$ by requiring
\begin{equation}
P(T_\mathrm{c_1})
<
10^{-5}.
\end{equation}
This criterion is used as a numerical proxy for the effective completion of the transition.

\begin{table}[!h]
    \centering
    \renewcommand{\arraystretch}{0.95} 
    \begin{tabular}{|wc{3cm}||wc{3cm}|wc{3cm}|wc{3cm}|}
    \hline\hline
    & \textbf{Set A} & \textbf{Set B} & \textbf{Set C} \\
    \hline\hline

    $\gamma$ & $4.1667\times10^{-8}$ & $4.1667 \times10^{-4}$ & $4.1667 \times10^{-4}$ \\ \hline
    $g$ & 1.05 & 1.05 & 1.05 \\ \hline
    $n_b$ & 20 & 20 & 0 \\ \hline
    $n_f$ & 20 & 20 & 20 \\ \hline
    \hline

    $V_0^{1/4}/{10^4 \TeV}$ & 0.71 & 70.71 & 70.71 \\ \hline
    $H_{\mathrm{TI}}/\TeV$ & $1.186 \times 10^{-8}$ & $1.186 \times 10^{-4}$ & $1.186 \times 10^{-4}$ \\ \hline
    $N_{\mathrm{TI}}$ & 4.6 & 11 & 11 \\ 
    \hline\hline

    $T_\mathrm{b}/\TeV$ & 242 & 112130 & 112130 \\ \hline
    $T_\mathrm{n}/\TeV$ & 1.243 & 1.189 & 1.548 \\ \hline
    $T_\mathrm{p}/\TeV$ & 1.22 & 1.164 & 1.519 \\ \hline
    $T_\mathrm{c_1}/\TeV$ & 1.214 & 1.158 & 1.511 \\ \hline
    $T_\mathrm{c_2}/\TeV$ & 1.152 & 1.152 & 1.504 \\ 
    \hline\hline
    \end{tabular}
    \caption{Benchmark parameter sets and the resulting thermal inflation and phase transition dynamics from \texttt{cosmoTransitions}. We take $y=y_b=y_f=1.09$ and $m = 1\TeV$.}
    \label{table:benchmarks}
\end{table}

\begin{figure}[!ht]
    \centering
    \includegraphics[width=1\linewidth]{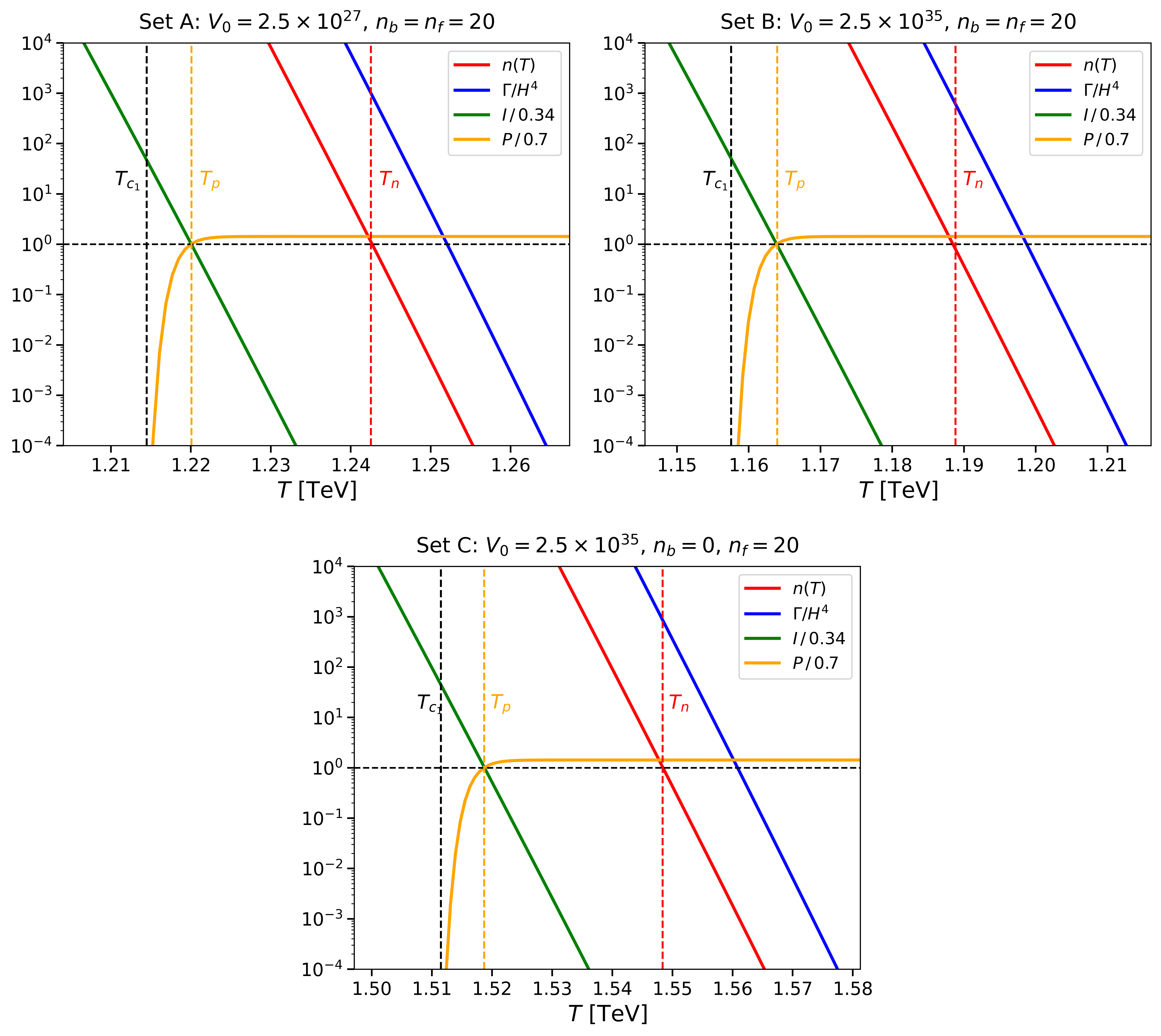}
    \caption{Evolution of the normalized nucleation and percolation diagnostics, $\Gamma/H^4$, $n(T)$, $I(T)/0.34$, and $P(T)/0.7$, near $T_\mathrm{c_2}$. The panels correspond to benchmark parameter sets A (top left), B (top right), and C (bottom center). The characteristic temperatures $T_\mathrm{n}$, $T_\mathrm{p}$, and $T_\mathrm{c_1}$ are indicated in each panel.}
    \label{fig:tunneling_set}
\end{figure}

\tref{table:benchmarks} summarizes the benchmark parameter sets and the resulting characteristic temperatures. For benchmark set A, the numerical value of $T_\mathrm{n}$ is consistent with the semi-analytic estimate obtained in the previous section. The numerical values of $T_\mathrm{c_2}$ also agree with the curvature-based estimates from \eq{eq:Tc2}. The temperature dependence of the quantities in \eqss{eq:tun_hub}{eq:nuc_temp}{eq:percol} is shown in \fref{fig:tunneling_set}.

Three qualitative features can be identified. First, the weakening of the potential barrier as the temperature decreases leads to a rapid increase in the nucleation rate. Second, the transition occurs at a temperature of order $m$ and completes within a narrow temperature interval. This behavior reflects the steep temperature dependence of the tunneling rate and its competition with the background expansion. Third, the small separation between $T_\mathrm{n}$ and $T_\mathrm{p}$ is consistent with the finite time required for the nucleated bubbles to grow and overlap in the expanding background.

\begin{figure}[!ht]
    \centering
    \includegraphics[width=1\linewidth]{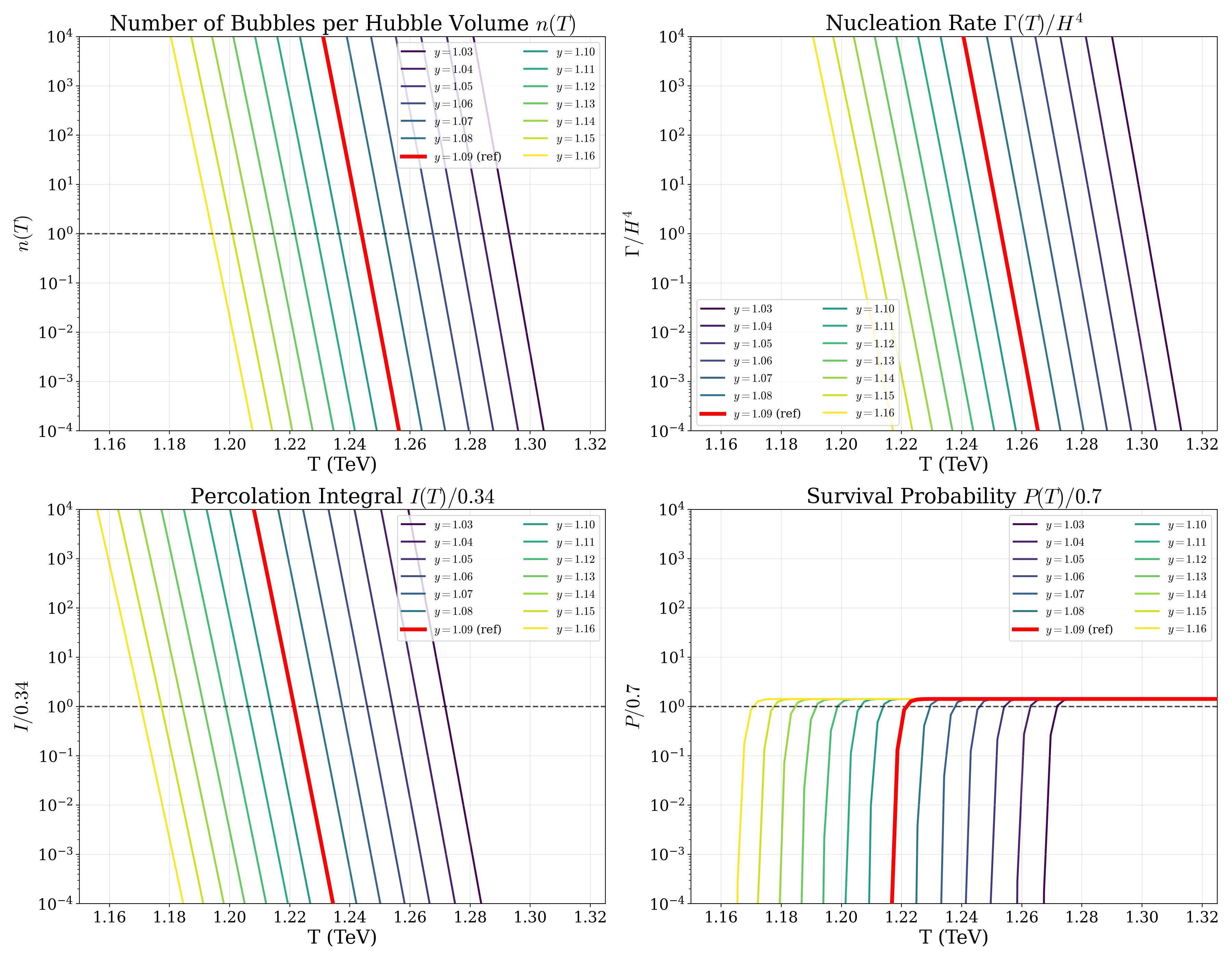}
    \caption{Thermal evolution of the bubble nucleation and percolation diagnostics for different values of the Yukawa coupling $y=y_b=y_f$. Shown clockwise from the top left are $n(T)$, $\Gamma/H^4$, $I(T)/0.34$, and $P(T)/0.7$. All other parameters are fixed to benchmark set A. The reference case $y=y_b=y_f=1.09$ is highlighted in red.}
    \label{fig:coupling_comparison_fermion_cosmo}
\end{figure}

\begin{figure}[!ht]
    \centering
    \includegraphics[width=1\linewidth]{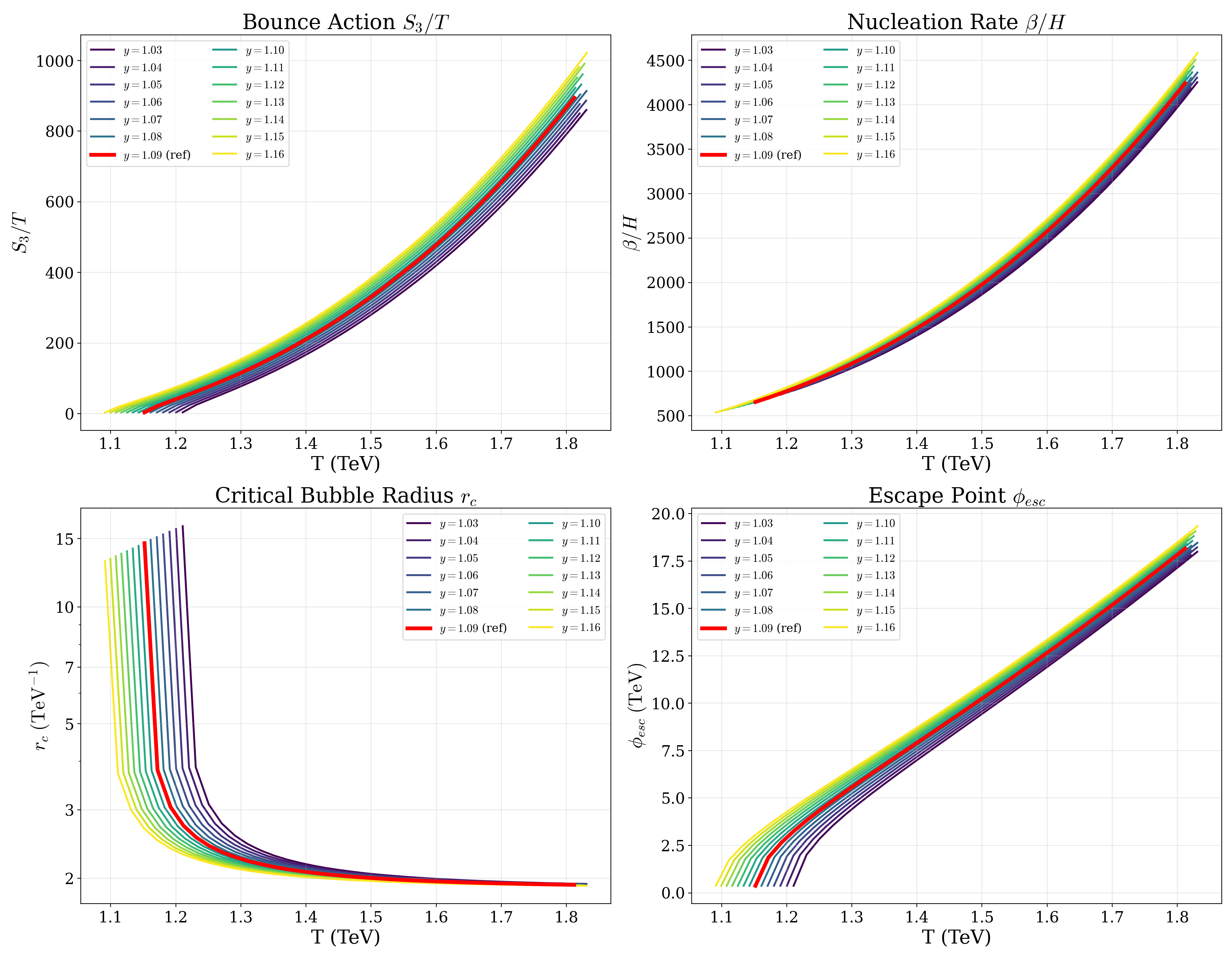}
    \caption{Thermal evolution of the critical-bubble parameters for different values of the Yukawa coupling $y=y_b=y_f$. Shown clockwise from the top left are $S_3/T$, $\beta/H$, the critical-bubble radius $r_\mathrm{c}$, and the escape point $\phi_\mathrm{e}$. All other parameters are fixed to benchmark set A. The reference case $y=y_b=y_f=1.09$ is highlighted in red.}
    \label{fig:coupling_comparison_fermion_rep5}
\end{figure}

We also scan the Yukawa coupling $y$ entering the thermal masses in \eq{eq:meff}. The quantities governing nucleation and percolation are shown in \fref{fig:coupling_comparison_fermion_cosmo}, while \fref{fig:coupling_comparison_fermion_rep5} presents $S_3/T$, the inverse duration parameter $\beta/H$, the critical-bubble radius $r_\mathrm{c}$, and the escape point $\phi_\mathrm{e}$. We define
\begin{equation}
\frac{\beta}{H}
=
T
\frac{d}{dT}
\left(
\frac{S_3}{T}
\right),
\label{eq:beta_over_H}
\end{equation}
evaluated at the percolation temperature unless stated otherwise.

Within the parameter range shown, increasing the Yukawa coupling strengthens the thermal contribution to the effective potential and tends to keep the flaton trapped near the origin down to a lower temperature. The transition is therefore shifted to a lower temperature. We also find $\beta/H\sim10^3$ around the percolation temperature. This large value indicates that the transition completes on a time scale much shorter than the Hubble time and tends to reduce the resulting GW amplitude. A quantitative assessment of the GW signal additionally requires the source efficiencies, the bubble wall dynamics, and the post-transition cosmological redshifting, which will be discussed in the following section.

\subsection{Lattice simulation}\label{subsec:lattice_simulation}

To follow the real-time dynamics of the phase transition, we solve the Langevin equation on a cubic lattice. Although modern cosmological lattice packages \cite{Felder:2000hq,Huang:2011gf,Figueroa:2021yhd} provide robust frameworks for scalar-field evolution, incorporating both stochastic thermal noise and a continuously evolving temperature into these packages is not straightforward. We therefore build on the Langevin approaches of \cite{Hiramatsu:2014uta,Dutka:2025oqt}, which provide a transparent treatment of thermal stochastic dynamics but were formulated without a time-dependent temperature or the spatially resolved evolution required in our analysis. Their framework is more readily extended to include these two effects, and we generalize it accordingly to describe the simultaneous cooling and spatial evolution of the field on the lattice.
 
 We rewrite the second-order equation as a coupled first-order system for $\phi$ and its conjugate velocity and evolve the deterministic terms using a second-order Runge-Kutta scheme. The additive Gaussian noise is sampled independently at each lattice site and time step. The same stochastic realization is used in the predictor and corrector stages of a given time step.

We introduce the rescaled coordinates
\begin{equation}
\tilde{\mathbf{x}}
=
\mu\mathbf{x},
\qquad
\tilde{t}
=
\mu t,
\end{equation}
where we choose $\mu=m$. We keep the flaton field $\phi$ in dimensionful units and define
\begin{equation}
\tilde{\eta}
=
\frac{\eta}{\mu},
\qquad
\tilde{H}
=
\frac{H}{\mu},
\qquad
\tilde{\xi}
=
\frac{\xi}{\mu^2},
\qquad
\tilde{\nabla}
=
\frac{\nabla}{\mu}.
\label{eq:lattice_rescaling}
\end{equation}
With these definitions, the Langevin equation becomes
\begin{equation}
\frac{\partial^2\phi}{\partial\tilde{t}^2}
+
\left(
\tilde{\eta}
+
3\tilde{H}
\right)
\frac{\partial\phi}{\partial\tilde{t}}
=
\frac{1}{a^2}
\tilde{\nabla}^2\phi
-
\frac{1}{\mu^2}
\frac{\partial V(\phi,T)}{\partial\phi}
+
\tilde{\xi}(\tilde{\mathbf{x}},\tilde{t}) .
\label{eq:langevin_rescaled}
\end{equation}

The scale factor and the temperature are updated continuously according to
\begin{equation}
\frac{d a}{d\tilde{t}}
=
\tilde{H}a,
\qquad
T(\tilde{t})
=
\frac{T_\mathrm{ini}}{a(\tilde{t})},
\label{eq:lattice_background_update}
\end{equation}
with $a(\tilde{t}_\mathrm{ini})=1$. The Hubble rate is evaluated from the background evolution described in \eq{eq:H_TI_compact}. The scaling $T\propto a^{-1}$ assumes adiabatic cooling and approximately constant effective relativistic degrees of freedom over the narrow temperature interval relevant for the transition.

The spatial Laplacian is discretized using the standard nearest-neighbor finite-difference stencil,
\begin{align}
\left(
\tilde{\nabla}^2\phi
\right)_{i,j,k}
&=
\frac{1}{(\Delta\tilde{x})^2}
\Big[
\phi_{i+1,j,k}
+
\phi_{i-1,j,k}
+
\phi_{i,j+1,k}
+
\phi_{i,j-1,k}
\nonumber
\\
&\hspace{3.1cm}
+
\phi_{i,j,k+1}
+
\phi_{i,j,k-1}
-
6\phi_{i,j,k}
\Big] .
\label{eq:lattice_laplacian}
\end{align}
This seven-point stencil approximates the continuum Laplacian with an error of order $(\Delta\tilde{x})^2$.

The noise correlator in \eq{eq:noise_corr} becomes
\begin{align}
\left\langle
\tilde{\xi}(\tilde{\mathbf{x}},\tilde{t})
\tilde{\xi}(\tilde{\mathbf{x}}',\tilde{t}')
\right\rangle
&=
\frac{D(T)}{a^3}
\delta^{(3)}
\left(
\tilde{\mathbf{x}}
-
\tilde{\mathbf{x}}'
\right)
\delta
\left(
\tilde{t}
-
\tilde{t}'
\right).
\label{eq:noise_corr_rescaled}
\end{align}
On the lattice, this is discretized as
\begin{align}
\left\langle
\tilde{\xi}_{\mathbf{i},n}
\tilde{\xi}_{\mathbf{j},m}
\right\rangle
&=
\frac{D(T_n)}
{
a^3(T_n)\,
\Delta\tilde{t}\,
(\Delta\tilde{x})^3
}
\,
\delta_{\mathbf{i}\mathbf{j}}
\delta_{nm},
\label{eq:noise_corr_lattice}
\end{align}
where $\mathbf{i}=(i,j,k)$ labels a lattice site and $n,m$ label time steps. Accordingly, the noise variable generated at each lattice site and time step is
\begin{equation}
\tilde{\xi}_{\mathbf{i},n}
=
\sqrt{
\frac{D(T_n)}
{
a^3(T_n)\,
\Delta\tilde{t}\,
(\Delta\tilde{x})^3
}
}
\,
\mathcal{G}_{\mathbf{i},n},
\label{eq:noise_lattice}
\end{equation}
where $\mathcal{G}_{\mathbf{i},n}$ is sampled independently from a standard normal distribution.

\subsubsection{Numerical setup and initial conditions}

We impose periodic boundary conditions and perform the reference simulations on a cubic lattice with linear grid size $N=256$, corresponding to $N^3=256^3$ lattice sites. Since $\tilde{\mathbf{x}}=m\mathbf{x}$ and $\tilde{t}=mt$, the rescaled lattice spacings are dimensionless. We choose
\begin{equation}
\Delta\tilde{x}
=
1,
\qquad
\Delta\tilde{t}
=
0.1.
\label{eq:lattice_reference_spacing}
\end{equation}
For $m=1\TeV$, these values correspond to
\begin{equation}
\Delta x
=
1\TeV^{-1},
\qquad
\Delta t
=
0.1\TeV^{-1}.
\end{equation}
The rescaled comoving box size and its dimensionful counterpart are
\begin{equation}
\tilde{L}
=
N\Delta\tilde{x}
=
256,
\qquad
L
=
\frac{\tilde{L}}{m}
=
256\TeV^{-1}.
\label{eq:lattice_box_size}
\end{equation}
The physical box size evolves as
\begin{equation}
L_\mathrm{phys}(\tilde{t})
=
a(\tilde{t})L.
\end{equation}

The reference timestep satisfies the Courant--Friedrichs--Lewy condition \cite{1967IBMJ...11..215C} associated with the gradient term,
\begin{equation}
\Delta\tilde{t}
<
\frac{
a(\tilde{t})\Delta\tilde{x}
}{
\sqrt{3}
}.
\label{eq:lattice_CFL}
\end{equation}
Since $a(\tilde{t}_\mathrm{ini})=1$ and the scale factor subsequently increases, the most restrictive bound is imposed at the initial time. Our choice satisfies
\begin{equation}
\Delta\tilde{t}
=
0.1
<
\frac{1}{\sqrt{3}}.
\end{equation}
The CFL condition is a necessary stability criterion for the gradient sector. The timestep dependence of the full evolution, including the potential term and stochastic source, is tested numerically below.

We adopt a constant damping coefficient,
\begin{equation}
\eta
=
T_\mathrm{ini},
\qquad
D(T)
=
2\eta T(\tilde{t}).
\label{eq:lattice_damping_prescription}
\end{equation}
The simulations begin at
\begin{equation}
T_\mathrm{ini}
=
1.230\TeV
\qquad
\text{for set B},
\end{equation}
and
\begin{equation}
T_\mathrm{ini}
=
1.600\TeV
\qquad
\text{for set C}.
\end{equation}
At the initial time, the field velocity vanishes,
\begin{equation}
\left.
\frac{\partial\phi_{\mathbf{i}}}
{\partial\tilde{t}}
\right|_{\tilde{t}=\tilde{t}_\mathrm{ini}}
=
0.
\end{equation}
The initial value of the field at each lattice site is sampled
independently from a Gaussian distribution,
\begin{equation}
\phi_{\mathbf{i}}(\tilde{t}_\mathrm{ini})
\sim
\mathcal{N}
\left(
0,
\sigma_\phi^2
\right),
\qquad
\sigma_\phi^2
=
10^{-10}\TeV^2,
\label{eq:lattice_initial_field}
\end{equation}
corresponding to
\begin{equation}
\sigma_\phi
=
10^{-5}\TeV.
\end{equation}
These initial fluctuations are site-local random seeds rather than pre-imposed bubble profiles. The subsequent nucleation and growth are generated dynamically by the Langevin evolution.

\begin{table}[!ht]
\centering
\begin{tabular}{lcc}
\hline
Quantity
&
Set B
&
Set C
\\
\hline
$N^3$
&
$256^3$
&
$256^3$
\\
$\Delta\tilde{x}$
&
$1$
&
$1$
\\
$\Delta\tilde{t}$
&
$0.1$
&
$0.1$
\\
$\tilde{L}=N\Delta\tilde{x}$
&
$256$
&
$256$
\\
$T_\mathrm{ini}/\TeV$
&
$1.230$
&
$1.600$
\\
$\eta/\TeV$
&
$1.230$
&
$1.600$
\\
$\sigma_\phi/\TeV$
&
$10^{-5}$
&
$10^{-5}$
\\
\hline
\end{tabular}
\caption{Reference lattice parameters and initial conditions. The
rescaled spacings $\Delta\tilde{x}$ and $\Delta\tilde{t}$ are
dimensionless. For $m=1\TeV$, they correspond to
$\Delta x=1\TeV^{-1}$ and $\Delta t=0.1\TeV^{-1}$.}
\label{table:lattice_setup}
\end{table}

\subsubsection{False-vacuum fraction and transition completion}

To quantify the progress of the transition, we classify each lattice site according to its local field value. We define a site as remaining in the false-vacuum region if
\begin{equation}
\left|
\phi_{\mathbf{i}}(\tilde{t})
\right|
<
\phi_\mathrm{cut}.
\label{eq:false_vacuum_site}
\end{equation}
The threshold is chosen with reference to the tunneling escape-point scale obtained from the bounce calculation. Near the nucleation temperature,
\begin{equation}
\phi_\mathrm{e}
\simeq
5\TeV,
\end{equation}
as shown in \fref{fig:coupling_comparison_fermion_rep5}. We conservatively
choose
\begin{equation}
\phi_\mathrm{cut}
=
2\phi_\mathrm{e}
=
10\TeV.
\label{eq:phi_cut}
\end{equation}
Sites satisfying
\begin{equation}
\left|
\phi_{\mathbf{i}}(\tilde{t})
\right|
\geq
\phi_\mathrm{cut}
\end{equation}
are classified as having left the false-vacuum basin and entered one of the two true-vacuum branches. This operational definition counts field configurations that have rolled sufficiently far away from the barrier; it does not require that the local field value has already relaxed to the zero-temperature vacuum expectation value.

The false-vacuum volume fraction of the lattice is
\begin{equation}
P_\mathrm{lat}(\tilde{t})
= 
\frac{1}{N^3}
\sum_{\mathbf{i}}
\Theta
\left(
\phi_\mathrm{cut}
-
\left|
\phi_{\mathbf{i}}(\tilde{t})
\right|
\right).
\label{eq:P_lat}
\end{equation}
We define the lattice transition-completion temperature by
\begin{equation}
P_\mathrm{lat}(T_\mathrm{c_1})
<
10^{-5}.
\label{eq:Tc1_lat}
\end{equation}

\begin{figure}[!ht]
    \centering
    \includegraphics[width=1\linewidth]
    {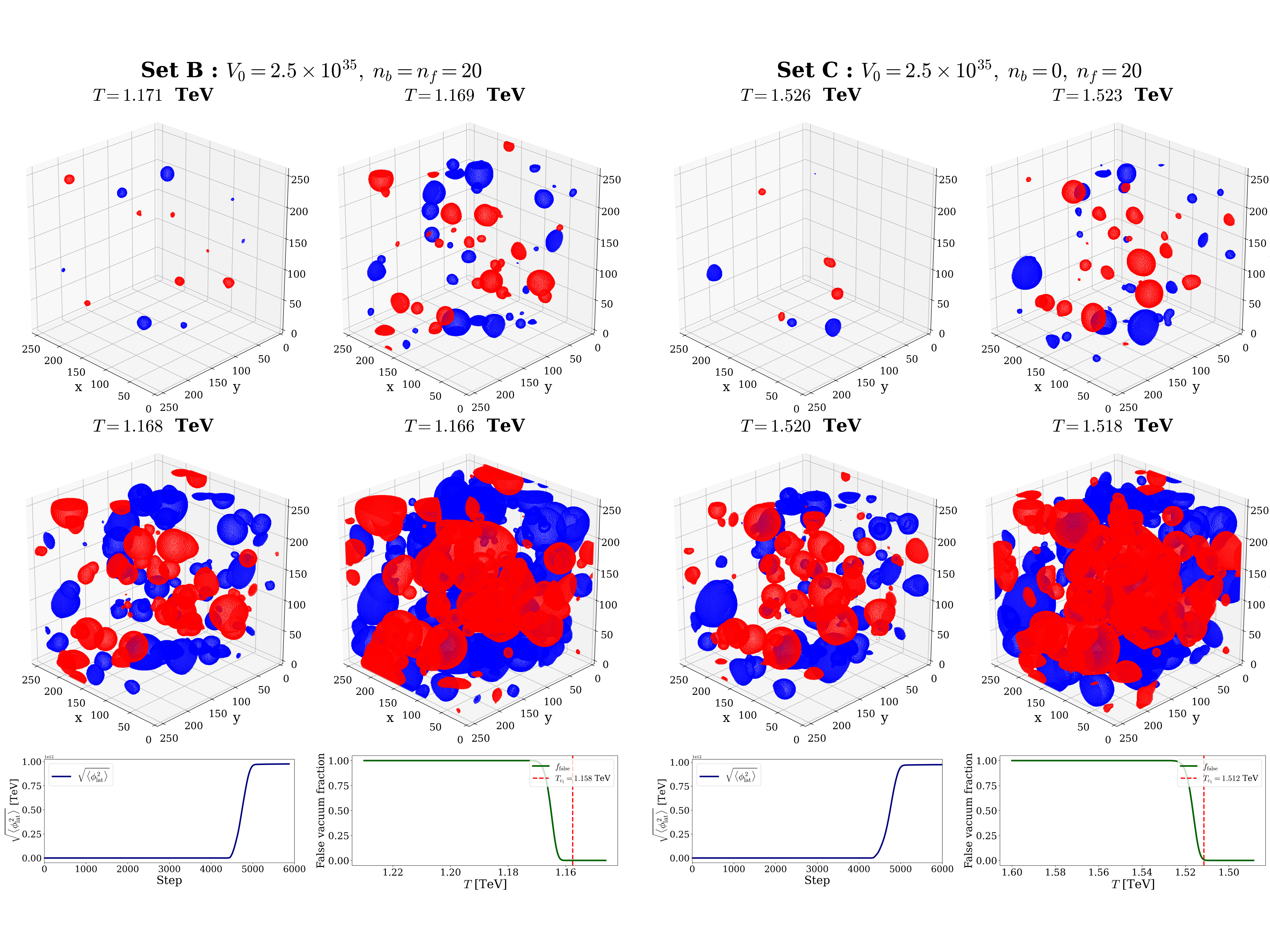}
    \caption{Evolution of the lattice field configurations for benchmark
    sets B (left column) and C (right column), with
    $y=y_b=y_f=1.09$. The top and middle rows show three-dimensional
    snapshots of regions that have left the false-vacuum basin:
    \protect\textcolor{red}{$\bullet$} indicates
    $\phi\geq\phi_\mathrm{cut}=10\TeV$, while
    \protect\textcolor{blue}{$\bullet$} indicates
    $\phi\leq-\phi_\mathrm{cut}=-10\TeV$. The bottom panels show the
    root-mean-square field amplitude and the lattice false-vacuum fraction
    $P_\mathrm{lat}$.}
    \label{fig:lattice_BC}
\end{figure}
\fref{fig:lattice_BC} illustrates the evolution of the field configurations and the volume-averaged root-mean-square amplitude. The lattice simulations give
\begin{equation}
T_\mathrm{c_1}
\simeq
1.160\TeV
\qquad
\text{for set B},
\end{equation}
and
\begin{equation}
T_\mathrm{c_1}
\simeq
1.509\TeV
\qquad
\text{for set C}.
\end{equation}
These values are close to the corresponding estimates obtained from the numerical bounce-action analysis in \tref{table:benchmarks}, namely $1.158\TeV$ and $1.511\TeV$, respectively. The characteristic initial bubble size observed in the reference simulations is also compatible, within the lattice resolution, with the critical radius $r_\mathrm{c}\sim3\TeV^{-1}$ inferred from the  bounce calculation; see
\fref{fig:coupling_comparison_fermion_rep5}.

\subsubsection{Convergence and robustness tests}

The white-noise source in \eq{eq:noise_lattice} is regulated by the lattice spacing. Accordingly, the simulations should be interpreted as an effective stochastic description with a finite ultraviolet cutoff. We do not attempt a continuum extrapolation in this work. Instead, we test whether the transition-completion temperature and the qualitative nucleation pattern are stable under moderate variations of the timestep, the spatial resolution, the box size, the random seed and the classification threshold.

We use the reference simulation in \tref{table:lattice_setup} and perform the convergence tests summarized in \tref{table:lattice_convergence}. The finer-resolution run keeps the comoving box size fixed while reducing $\Delta\tilde{x}$. The larger-volume run tests sensitivity to periodic boundary conditions.

\begin{table*}[!ht]
\centering
\begin{tabular}{lcccccc}
\hline
Run
&
$N$
&
$\Delta\tilde{x}$
&
$\Delta\tilde{t}$
&
$\tilde{L}$
&
$T_\mathrm{c_1}^{(B)}/\TeV$
&
$T_\mathrm{c_1}^{(C)}/\TeV$
\\
\hline
Reference
&
$256$
&
$1.0$
&
$0.10$
&
$256$
&
$1.160$
&
$1.509$
\\
Smaller timestep
&
$256$
&
$1.0$
&
$0.05$
&
$256$
&
$1.162$
&
$1.510$
\\
Finer spatial resolution
&
$512$
&
$0.5$
&
$0.05$
&
$256$
&
$1.182$
&
$1.537$
\\
Larger comoving volume
&
$384$
&
$1.0$
&
$0.10$
&
$384$
&
$1.159$
&
$1.508$
\\
\hline
\end{tabular}
\caption{Convergence tests for the lattice simulations.}
\label{table:lattice_convergence}
\end{table*}

The completion temperatures obtained from the convergence runs differ from the reference values by at most $1.90\%$ for set B and $1.86\%$ for set C. The localized nucleation events and the subsequent growth and overlap of the transitioned regions remain visible in all runs. In particular, the qualitative distinction between localized bubble growth and an immediate global phase-mixing instability is unchanged.

We also test the sensitivity to the $\phi_{\mathrm{cut}}$ threshold used in \eq{eq:P_lat}. The results are summarized in \tref{table:lattice_threshold_test}. Varying $\phi_\mathrm{cut}$ between $1.5\phi_\mathrm{e}$ and $2.5\phi_\mathrm{e}$ changes the extracted completion temperature by less than $0.07\%$.

\begin{table}[!ht]
\centering
\begin{tabular}{ccc}
\hline
$\phi_\mathrm{cut}$
&
$T_\mathrm{c_1}^{(B)}/\TeV$
&
$T_\mathrm{c_1}^{(C)}/\TeV$
\\
\hline
$1.5\phi_\mathrm{e}=7.5\TeV$
&
$1.160$
&
$1.509$
\\
$2.0\phi_\mathrm{e}=10.0\TeV$
&
$1.160$
&
$1.509$
\\
$2.5\phi_\mathrm{e}=12.5\TeV$
&
$1.160$
&
$1.508$
\\
\hline
\end{tabular}
\caption{Sensitivity of the lattice transition-completion temperature to the threshold used to define the false-vacuum fraction.}
\label{table:lattice_threshold_test}
\end{table}

\subsubsection{Role of the evolving cosmological background}

To isolate the origin of the difference from the phase-mixing picture reported in \cite{Hiramatsu:2014uta}, we perform diagnostic control simulations in which the background treatment is varied while the microphysical parameters are held fixed. These control runs are intended to identify the numerical ingredient responsible for the difference; they should not be interpreted as independent cosmological models. We compare:
\begin{itemize}
    \item a fixed-background run with $a=1$ and $T=T_\mathrm{ini}$;
    \item a prescribed-cooling run in which $T(\tilde{t})$ follows the same
    history as in the full simulation while the explicit expansion terms
    in the Langevin equation are switched off; and
    \item the full expanding-background run in which the Hubble friction,
    the gradient suppression, the noise dilution, and the temperature
    evolution are included consistently.
\end{itemize}

\begin{table*}[!ht]
\centering
\begin{tabular}{lcccc}
\hline
Background treatment
&
$3H\dot{\phi}$
&
$a^{-2}\nabla^2\phi$
&
$a^{-3}$ noise dilution
&
Observed transition pattern
\\
\hline
Fixed background
&
No
&
No expansion
&
No
&
Global thermal fluctuation
\\
Prescribed cooling only
&
No
&
No expansion
&
No
&
Localized nucleation and growth
\\
Full expanding background
&
Yes
&
Yes
&
Yes
&
Localized nucleation and growth
\\
\hline
\end{tabular}
\caption{Diagnostic control simulations used to isolate the effect of the
evolving cosmological background.}
\label{table:lattice_background_controls}
\end{table*}

The control simulations show that global thermal fluctuations only occurs at fixed temperature. This comparison indicates that the temperature drop is an important ingredient in recovering the bubble nucleation picture in the parameter region considered here.

For benchmark sets B and C, the lattice simulations therefore support a transition picture dominated by localized nucleation and subsequent growth of true-vacuum regions rather than by an immediate global phase-mixing instability. This conclusion is restricted to the parameter region and the effective stochastic framework studied in this work. The convergence, threshold-sensitivity and background-control tests provide evidence that the result is not an artifact of a particular lattice discretization or operational definition.

\section{Gravitational waves from bubbles}\label{sec:gwb}

We estimate the stochastic gravitational-wave (GW) background generated by
the first-order phase transition at the end of thermal inflation. In this
section, we focus on the parameter family
\begin{equation}
n_b=20,
\qquad
n_f=20,
\qquad
y=y_b=y_f=1.09,
\qquad
g=1.05,
\qquad
T_\mathrm{d}=0.1\TeV ,
\label{eq:GW_fixed_parameters}
\end{equation}
and set $m=1\TeV$. The benchmark value $T_\mathrm{d}=0.1\TeV$ is motivated by
\cite{Easther:2008sx}. For the illustrative spectra shown below, we use
\begin{equation}
\frac{\beta}{H_*}=1000,
\qquad
v_\mathrm{w}=1,
\qquad
g_{*,\mathrm{d}}=g_{*s,\mathrm{d}}=100,
\qquad
h=0.678 ,
\label{eq:GW_benchmark_assumptions}
\end{equation}
where $H_*\simeq H_\mathrm{TI}$ denotes the Hubble rate at GW production.
The dependence on $\gamma$ is displayed explicitly.

The relative importance of the GW sources depends on the bubble wall
dynamics. Particles coupled to the flaton acquire large masses inside the
true-vacuum bubble. Their decay after crossing the wall does not remove the
momentum already transferred to the wall. However, the usual leading-order
friction estimate assumes that these species remain sufficiently populated
in the false-vacuum plasma. This assumption is model dependent, because it
depends on their distribution functions and reaction rates. We therefore
use two limiting descriptions to bracket the theoretical uncertainty:
\begin{itemize}
    \item a plasma-coupled benchmark, in which sound waves and turbulence
    provide the dominant GW sources; and
    \item a weak-friction limiting case, in which bubble wall
    collisions can become important.
\end{itemize}
These two limits should not be interpreted as simultaneously saturated
source contributions.

\subsection{Bubble wall friction and source uncertainty}

At leading order, the frictional pressure exerted by the plasma on an
ultra-relativistic bubble wall is estimated as \cite{Bodeker:2009qy}
\begin{equation}
\mathcal{P}_{\mathrm{LO}}(T)
=
\frac{T^2}{48}
\left(
2\sum_i n_{b_i}\Delta m_{b_i}^2
+
\sum_j n_{f_j}\Delta m_{f_j}^2
\right),
\label{eq:friction_LO}
\end{equation}
where $n_{b_i}$ and $n_{f_j}$ denote the bosonic and fermionic degrees of
freedom, respectively.

To make the population assumption explicit, we introduce effective
occupation factors $0\leq\chi_i(T)\leq1$:
\begin{equation}
\mathcal{P}_{\mathrm{LO}}(T)
=
\frac{T^2}{48}
\left[
2\sum_i
\chi_{b_i}(T)n_{b_i}\Delta m_{b_i}^2
+
\sum_j
\chi_{f_j}(T)n_{f_j}\Delta m_{f_j}^2
\right].
\label{eq:friction_LO_weighted}
\end{equation}
The equilibrium estimate corresponds to $\chi_i=1$, while suppressed
abundances or inefficient replenishment in the false-vacuum plasma reduce
the effective pressure.

Using \eq{eq:meff}, the mass-squared differences across the wall are
\begin{equation}
\Delta m_{b_i}^2
=
\frac{1}{2}y_{b_i}^2\phi_0^2,
\qquad
\Delta m_{f_j}^2
=
\frac{1}{2}y_{f_j}^2\phi_0^2 .
\label{eq:mass_jump_wall}
\end{equation}
Approximating the driving pressure by
$\Delta V(T_\mathrm{n})\simeq V_0=m^2\phi_0^2/4$, the equilibrium estimate
becomes
\begin{equation}
\frac{\Delta V(T_\mathrm{n})}
{\mathcal{P}_{\mathrm{LO}}(T_\mathrm{n})}
\simeq
\frac{24m^2}
{
T_\mathrm{n}^2
\left(
2\sum_i n_{b_i}y_{b_i}^2
+
\sum_j n_{f_j}y_{f_j}^2
\right)
}.
\label{eq:runaway_ratio_general}
\end{equation}
For $y_b=y_f=y$, this reduces to
\begin{equation}
\frac{\Delta V(T_\mathrm{n})}
{\mathcal{P}_{\mathrm{LO}}(T_\mathrm{n})}
\simeq
\frac{24}
{y^2(2n_b+n_f)}
\left(
\frac{m}{T_\mathrm{n}}
\right)^2 .
\label{eq:runaway_ratio_common}
\end{equation}
Note that our estimate of the bubble-wall friction does not account for particle decays within the wall profile. Since the nucleated bubbles are expected to lie in the thick-wall regime, particles in the thermal plasma may decay before traversing the entire wall and reaching the true-vacuum region, where $\phi=\phi_0$. If their decay rates are sufficiently small, however, they continue to gain mass as they propagate toward the global minimum. A quantitative treatment of these non-equilibrium wall--plasma dynamics requires dedicated real-time kinetic simulations and is beyond the scope of the present work.

For the equilibrium population assumption and $T_\mathrm{n}\sim m$, the
ratio in \eq{eq:runaway_ratio_common} is smaller than unity. The
leading-order criterion then does not favor a runaway wall. This conclusion
is conditional on the false-vacuum particle populations. If the effective
occupation factors in \eq{eq:friction_LO_weighted} are strongly suppressed,
the bubble wall collision contribution may become relevant. Moreover,
transition radiation can limit the acceleration even when the leading-order
friction is small \cite{Bodeker:2017cim}. A microscopic determination of
$v_\mathrm{w}$ and the source efficiencies requires a kinetic treatment of
the plasma and is beyond the scope of this work.

We therefore write
\begin{equation}
\Omega_{\mathrm{GW},0}(f)h^2
=
\Omega_{\phi,0}(f)h^2
+
\Omega_{\mathrm{sw},0}(f)h^2
+
\Omega_{\mathrm{turb},0}(f)h^2 ,
\label{eq:Omega_total_GW}
\end{equation}
where $\Omega_{\phi,0}$ denotes the bubble wall collision contribution.
For the plasma-coupled benchmark, we set $\Omega_{\phi,0}\simeq0$. For the
weak-friction limiting case, we retain $\Omega_{\phi,0}$ as an optimistic
upper-envelope estimate.

\subsection{Redshift through the flaton matter-dominated era}

After the phase transition, coherent flaton oscillations behave as
non-relativistic matter until the flaton decays. This intermediate
matter-dominated era suppresses the GW energy-density fraction and shifts
the characteristic frequencies. We define
\begin{equation}
R_\mathrm{md}
\equiv
\frac{a_*}{a_\mathrm{d}}
=
\left[
\frac{\rho_\mathrm{r}(T_\mathrm{d})}
{V_0}
\right]^{1/3}
=
\left[
\frac{\pi^2g_{*,\mathrm{d}}T_\mathrm{d}^4}
{30V_0}
\right]^{1/3},
\label{eq:Rmd}
\end{equation}
where $a_*$ denotes the scale factor at GW production and $a_\mathrm{d}$
denotes the scale factor at flaton decay.

For $g_{*,\mathrm{d}}=100$, this becomes
\begin{equation}
R_\mathrm{md}
\simeq
6.90\times10^{-7}
\left(
\frac{T_\mathrm{d}}{0.1\TeV}
\right)^{4/3}
\left(
\frac{V_0^{1/4}}{10^4\TeV}
\right)^{-4/3}.
\label{eq:Rmd_numeric}
\end{equation}

Entropy conservation after flaton decay gives
\begin{equation}
\frac{a_\mathrm{d}}{a_0}
=
\left(
\frac{g_{*s,0}}
{g_{*s,\mathrm{d}}}
\right)^{1/3}
\frac{T_0}{T_\mathrm{d}} .
\label{eq:ad_a0}
\end{equation}
It is useful to define the redshifted Hubble frequency
\begin{equation}
f_{H,0}
\equiv
H_*R_\mathrm{md}
\frac{a_\mathrm{d}}{a_0}.
\label{eq:fH0}
\end{equation}
Using $g_{*s,0}=3.91$, $T_0=2.7255\,\mathrm{K}$, and
\eq{eq:GW_benchmark_assumptions}, we obtain
\begin{equation}
f_{H,0}
\simeq
2.01\times10^{-2}\,\mathrm{Hz}
\left(
\frac{V_0^{1/4}}{10^4\TeV}
\right)^{2/3}
\left(
\frac{T_\mathrm{d}}{0.1\TeV}
\right)^{1/3}.
\label{eq:fH0_numeric}
\end{equation}

\subsection{GW source spectra}

For the weak-friction limiting case, we follow the envelope-approximation for bubble wall collisions \cite{Caprini:2015zlo,Weir:2016tov}:
\begin{align}
\Omega_{\phi,0}(f)h^2
&=
1.67\times10^{-5}
\left(
\frac{H_*}{\beta}
\right)^2
\left(
\frac{\kappa_\phi\alpha_*}
{1+\alpha_*}
\right)^2
\left(
\frac{100}{g_{*,\mathrm{d}}}
\right)^{1/3}
\nonumber
\\
&\quad\times
\left(
\frac{0.11v_\mathrm{w}^3}
{0.42+v_\mathrm{w}^2}
\right)
R_\mathrm{md}
\fn{S_\phi}{\frac{f}{f_{\phi,0}}},
\nonumber
\\
& \simeq 8.93 \times 10^{-19} \left(\dfrac{\beta/H_*}{1000}\right)^{-2}\left(\dfrac{V_0^{1/4}}{10^4\TeV}\right)^{-4/3}\left(\dfrac{T_d}{0.1\TeV}\right)^{4/3}S_{\phi}\left(\dfrac{f}{f_{\phi,0 }}\right)\label{eq:Omega_phi}
\\
\fn{S_\phi}{x}
&=
\frac{3.8x^{2.8}}
{1+2.8x^{3.8}},
\label{eq:S_phi}
\\
f_{\phi,0}
&=
\frac{0.62}
{1.8-0.1v_\mathrm{w}+v_\mathrm{w}^2}
\frac{\beta}{H_*}
f_{H,0}
\nonumber
\\
&\simeq4.62\mathrm{Hz}\left(\dfrac{\beta / H_*}{1000}\right)\left(\dfrac{V_0^{1/4}}{10^4\TeV}\right)^{2/3}\left(\dfrac{T_d}{0.1\TeV}\right)^{1/3}
\label{eq:f_phi_peak}
\end{align}
Here $\kappa_\phi$ denotes the fraction of released vacuum energy stored in
scalar-field gradients. The choice $\kappa_\phi=1$ and $\alpha_*\gg1$
defines an optimistic upper envelope.

For the sound-wave contribution, we use the simulation-based fit of
\cite{Hindmarsh:2017gnf}, including its corrected normalization:
\begin{align}
\Omega_{\mathrm{sw},0}(f)h^2
&=
2.061\,
h^2F_{\mathrm{gw},0}\,
\Gamma_{\mathrm{ad}}^2
\bar{U}_f^4
\left(
H_*R_*
\right)
\widetilde{\Omega}_{\mathrm{gw}}
R_\mathrm{md}
\Upsilon_\mathrm{sw}
\fn{S_\mathrm{sw}}{\frac{f}{f_{\mathrm{sw},0}}},
\label{eq:Omega_sw}
\\
\fn{S_\mathrm{sw}}{x}
&=
x^3
\left(
\frac{7}{4+3x^2}
\right)^{7/2}.
\label{eq:S_sw}
\end{align}
Here
\begin{equation}
h^2F_{\mathrm{gw},0}
=
1.64\times10^{-5}
\left(
\frac{100}{g_{*,\mathrm{d}}}
\right)^{1/3},
\qquad
\Gamma_{\mathrm{ad}}
\simeq
\frac{4}{3},
\qquad
\widetilde{\Omega}_{\mathrm{gw}}
=
1.2\times10^{-2},
\label{eq:sound_fit_parameters}
\end{equation}
$\Gamma_{\mathrm{ad}}$ is an adiabatic index, $\bar{U}_f$ denotes the root-mean-square fluid velocity and $R_*$ is the bubble radius. In the
strong-transition benchmark, we take
\begin{equation}
\bar{U}_f
\simeq
\sqrt{\frac{3}{4}},
\qquad
R_*
\simeq
(8\pi)^{1/3}
\frac{v_\mathrm{w}}{\beta}.
\label{eq:sound_benchmark_parameters}
\end{equation}

If the bubble wall expands rapidly, the acoustic wave does not necessarily last for full Hubble time. The duration of the sound wave oscillation $\Upsilon_\mathrm{sw}$ is estimated by comparing the Hubble time and the characteristic time of bubble collision as
\begin{equation}
\Upsilon_\mathrm{sw}
\simeq
\min\left[
1,
H_*\tau_\mathrm{sw}
\right],
\qquad
\tau_\mathrm{sw}
\simeq
\frac{R_*}{\bar{U}_f},
\label{eq:sound_lifetime}
\end{equation}
following \cite{Ellis:2020awk}. For \eq{eq:GW_benchmark_assumptions},
\begin{equation}
\Upsilon_\mathrm{sw}
\simeq
3.38\times10^{-3}.
\label{eq:Upsilon_sw_numeric}
\end{equation}

Thus, \eq{eq:Omega_sw} is estimated as 
\begin{equation}
    \Omega_{\text{sw},0}(f)h^2 \simeq 2.76 \times 10^{-18} \left( \frac{\beta /H_*}{1000} \right)^{-2} \left( \frac{T_d}{0.1 \text{ TeV}} \right)^{4/3} \left( \frac{V_0^{1/4}}{10^4 \text{ TeV}} \right)^{-4/3} S_{\text{sw}}\left( \frac{f}{f_{\text{sw},0}} \right)
\end{equation}
and its peak frequency is

\begin{align}
f_{\mathrm{sw},0}
&\simeq
\frac{1.16}{v_\mathrm{w}}
\frac{\beta}{H_*}
f_{H,0}
\nonumber
\\
&\simeq 2.33 \times 10^{1} \text{ Hz} \left( \frac{\beta/H_*}{1000} \right) \left( \frac{V_0^{1/4}}{10^4 \text{ TeV}} \right)^{2/3} \left( \frac{T_d}{0.1 \text{ TeV}} \right)^{1/3}
\label{eq:f_sw_peak}
\end{align}

For the turbulence contribution, we use \cite{Caprini:2015zlo}
\begin{align}
\Omega_{\mathrm{turb},0}(f)h^2
&=
3.35\times10^{-4}
\left(
\frac{H_*}{\beta}
\right)
\left(
\frac{\kappa_\mathrm{turb}\alpha_*}
{1+\alpha_*}
\right)^{3/2}
\left(
\frac{100}{g_{*,\mathrm{d}}}
\right)^{1/3}
v_\mathrm{w}
R_\mathrm{md}
\fn{S_\mathrm{turb}}{f},
\nonumber
\\
&\simeq 2.58 \times 10^{-15} \left( \frac{\beta/H_*}{1000} \right)^{-1} \left( \frac{T_d}{0.1 \text{ TeV}} \right)^{4/3} \left( \frac{V_0^{1/4}}{10^4 \text{ TeV}} \right)^{-4/3} S_{\text{turb}}(f)
\label{eq:Omega_turb}
\\
\fn{S_\mathrm{turb}}{f}
&=
\frac{
\left(
f/f_{\mathrm{turb},0}
\right)^3
}{
\left(
1+f/f_{\mathrm{turb},0}
\right)^{11/3}
\left(
1+8\pi f/f_{H,0}
\right)
},
\label{eq:S_turb}
\end{align}
where
\begin{equation}
\kappa_\mathrm{turb}
=
\epsilon_\mathrm{turb}\kappa_\mathrm{v},
\qquad
\epsilon_\mathrm{turb}=0.05,
\qquad
\kappa_\mathrm{v}\simeq1.
\label{eq:kappa_turb}
\end{equation}
$\kappa_{\nu}$ denotes the fraction of released vacuum energy for sound-wave and $\epsilon_{\mathrm{turb}}$ is a fraction of turbulent motion to bulk motion. The peak frequency is estimated as
\begin{align}
f_{\mathrm{turb},0}
&\simeq
\frac{1.64}{v_\mathrm{w}}
\frac{\beta}{H_*}
f_{H,0}
\nonumber
\\
&\simeq 3.29 \times 10^{1} \text{ Hz} \left( \frac{\beta/H_*}{1000} \right) \left( \frac{V_0^{1/4}}{10^4 \text{ TeV}} \right)^{2/3} \left( \frac{T_d}{0.1 \text{ TeV}} \right)^{1/3}
\label{eq:f_turb_peak}
\end{align}
At the peak frequency, the spectral shape evaluates to $S_{\mathrm{turb}}(f_{\mathrm{turb},0}) = \frac{1}{2^{11/3}\left[1+8\pi\left(\frac{1.64}{v_w}\frac{\beta}{H_*}\right)\right]} \simeq 2 \times 10^{-6}$. This restricts the maxim um amplitude of \eq{eq:Omega_turb} to $\simeq 4.93 \times 10^{-21}$, making the turbulence contribution negligible in this regime.

\subsection{Numerical spectra and experimental sensitivity}

For the fixed parameter family in \eq{eq:GW_fixed_parameters}, the potential term can be expressed using \eq{eq:pot_params},
\begin{equation}
V_0^{1/4}
=
\left(
\frac{\gamma m\Mpl}{2}
\right)^{1/2}.
\label{eq:V04_gamma}
\end{equation}
Therefore, the leading dependence on $\gamma$ for the GW spectra and their peak frequency is
\begin{equation}
\Omega_{\mathrm{GW},0}^{\mathrm{peak}}
\propto
R_\mathrm{md}
\propto
\gamma^{-2/3},
\qquad
f_{\mathrm{peak}}
\propto
f_{H,0}
\propto
\gamma^{1/3}.
\label{eq:GW_gamma_scaling}
\end{equation}
Thus, decreasing $\gamma$ reduces the dilution during flaton matter domination, raises the GW amplitude, and shifts the characteristic frequencies toward lower values.

\begin{figure*}[t]
    \centering
    \includegraphics[width=1\textwidth]
    {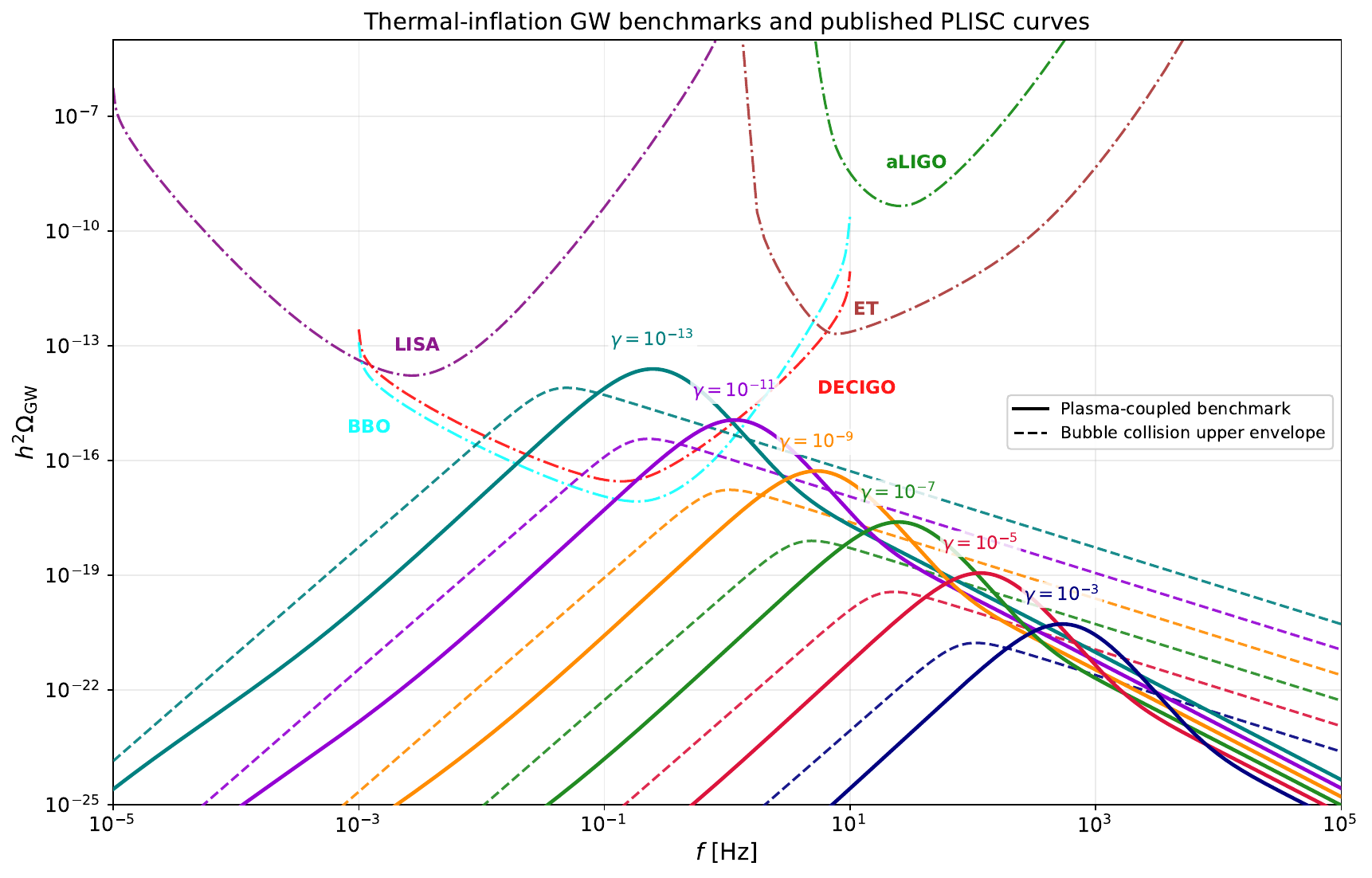}
    \caption{Present-day GW spectra for
    $n_b=n_f=20$, $y=y_b=y_f=1.09$, $g=1.05$,
    $T_\mathrm{d}=0.1\TeV$, $\beta/H_*=1000$, and $v_\mathrm{w}=1$ depending on $\gamma$.
    \rule[0.5ex]{0.5cm}{0.8pt}: plasma-coupled benchmark $\Omega_{\mathrm{sw},0}+\Omega_{\mathrm{turb},0}$
    \hdashrule[0.5ex]{1cm}{1pt}{1ex 0.5ex}: Upper envelope of bubble collisions in the weak-friction limit $\Omega_{\phi,0}$ with $\kappa_\phi=1$ and $\alpha_*\gg1$. The two source scenarios bracket different assumptions about the
    wall--plasma dynamics and should not be interpreted as simultaneously
    saturated contributions.}
    \label{fig:GW_gamma_updated}
\end{figure*}

\fref{fig:GW_gamma_updated} compares the two limiting source scenarios.
Solid curves show the plasma-coupled benchmark,
$\Omega_{\mathrm{sw},0}+\Omega_{\mathrm{turb},0}$, including the finite acoustic-lifetime suppression. Dashed curves show the upper envelope of bubble collisions with $\kappa_\phi=1$ and $\alpha_*\gg1$. The detector curves are the published one-year, threshold-SNR$=1$ power-law-integrated sensitivity curves for LISA, DECIGO,\ BBO, ET, and the aLIGO Hanford--Livingston network \cite{Giblin:2014gra, Schmitz:2020syl}. The GW spectra of both scenarios meet the sensitivity curves for future detectors where $\gamma\lesssim10^{-11}$.

The spectra in \fref{fig:GW_gamma_updated} should be interpreted as a range
rather than as a unique prediction. A more precise determination requires
a microscopic calculation of the false-vacuum particle distributions,
reaction rates, wall velocity, and the energy partition among scalar
gradients, bulk fluid motion, and turbulence. The detector comparison
therefore indicates the regions that can become phenomenologically
interesting under the corresponding limiting assumptions, rather than a
model-independent detectability threshold. It has been recently pointed out that for strong phase transitions, the GW spectrum generated by post-collision bulk-fluid motion can closely resemble the spectrum from bubble wall collisions \cite{Lewicki:2022pdb,Lewicki:2025hxg}. We leave the detailed investigation of these post-collision dynamics, which could potentially unify our two limiting source scenarios, for future work.

For values of $\beta/H_*$ different from $1000$, the characteristic
frequencies scale as
\begin{equation}
f_{\phi,0},
\,
f_{\mathrm{sw},0},
\,
f_{\mathrm{turb},0}
\propto
\frac{\beta}{H_*}.
\end{equation}
The scalar-collision upper envelope scales as
\begin{equation}
\Omega_{\phi,0}^{\mathrm{peak}}
\propto
\left(
\frac{\beta}{H_*}
\right)^{-2},
\end{equation}
while
\begin{equation}
\left.
\Omega_{\mathrm{sw},0}^{\mathrm{peak}}
\right|_{\Upsilon_\mathrm{sw}=1},
\,
\Omega_{\mathrm{turb},0}^{\mathrm{peak}}
\propto
\left(
\frac{\beta}{H_*}
\right)^{-1}.
\label{eq:GW_beta_scaling}
\end{equation}
In the short-lived acoustic regime,
$\Upsilon_\mathrm{sw}\propto(\beta/H_*)^{-1}$, so that
\begin{equation}
\Omega_{\mathrm{sw},0}^{\mathrm{peak}}
\propto
\left(
\frac{\beta}{H_*}
\right)^{-2}.
\label{eq:GW_beta_scaling_short_lived}
\end{equation}

\section{Conclusion}\label{sec:conclusion}

In this work, we investigated the first-order phase transition that terminates thermal inflation and estimated the associated stochastic gravitational-wave (GW) signals. Starting from the finite-temperature effective potential of the flaton, we obtained semi-analytic estimates of the characteristic transition temperatures and the three-dimensional Euclidean bounce action. The resulting nucleation temperature agrees with an independent numerical calculation performed using the \texttt{CosmoTransitions} package \cite{Wainwright:2011kj}. We then evolved the flaton field using a Langevin equation on an expanding three-dimensional lattice, incorporating the Hubble expansion and the temperature evolution of the effective potential throughout the simulation.

For the benchmark parameter sets considered in \tref{table:benchmarks}, the lattice evolution exhibits localized nucleation of true-vacuum bubbles, followed by their expansion, overlap, and eventual completion of the transition. The transition-completion temperatures are consistent with the bounce-action estimates. These results support a picture based on bubble nucleation and percolation in the parameter region investigated here. This differs from the conclusion of \cite{Hiramatsu:2014uta}, where the transition was found to proceed through phase mixing instead of bubble formation, implying the absence of an appreciable bubble-generated GW signal. The discrepancy may be associated with the treatment of the cosmological background, including the continuous Hubble expansion and the temperature evolution of the effective potential during the real-time lattice evolution. A systematic comparison of the two numerical setups is required to identify its precise origin. Our conclusion applies to the parameter region and dynamical assumptions explored in this work and should not be interpreted as a universal statement about all realizations of thermal inflation.

We also estimated the GW spectrum generated during the transition. The intermediate flaton-dominated matter era dilutes the GW energy density and shifts the characteristic frequencies, leading to the scaling relations
\begin{equation}
\Omega_{\mathrm{GW},0}^{\mathrm{peak}}
\propto
\gamma^{-2/3},
\qquad
f_{\mathrm{peak}}
\propto
\gamma^{1/3}.
\end{equation}
The normalization of the spectrum remains uncertain because it depends sensitively on the wall--plasma dynamics. Our leading-order estimate of the plasma friction assumes that the particle species coupled to the flaton remain sufficiently abundant in the false-vacuum plasma. If the friction is sizable, the sound-wave and magnetohydrodynamic-turbulence contributions provide an appropriate plasma-coupled benchmark. In the weak-friction regime, the scalar-field collision contribution may become important. Additional uncertainties arise from transition radiation and from the finite lifetime of the acoustic source \cite{Bodeker:2009qy,Bodeker:2017cim,Ellis:2020awk}. We therefore present an indicative range bounded by a plasma-coupled spectrum and an upper envelope from bubble collisions in the weak-friction limit, instead of a unique prediction. A comparison with published power-law-integrated sensitivity curves \cite{Schmitz:2020syl} shows that decreasing $\gamma$ shifts the signal toward the projected sensitivity ranges of future interferometers, although the precise detectability threshold remains model dependent.

A microscopic kinetic treatment of the particle distribution functions, reaction rates, wall velocity, and efficiency factors for the transfer of vacuum energy into scalar-field gradients, bulk fluid motion, and turbulence would reduce the dominant uncertainty in the GW prediction. The lattice analysis could also be generalized to a complex flaton field. This would permit a study of axionic-string formation and evolution during the transition and clarify whether these defects leave an additional observable signature of thermal inflation \cite{Maji:2023fhv,Jeong:2023iei}.

\acknowledgements
This work is supported by T\"{U}B\.{I}TAK-ARDEB-1001 program under project 123F257. The authors thank Ewan Stewart, Enes Zeybek, Gürcan Coşgel, Saya Mahmoudi, Atakan Tuncel,  Alpcan Bahar, Asra Frat, Furkan Biber, Alp Özmen, Adem Kılıç, Ahmet Şekerci, and Tae Hyun Jung for the helpful discussions.

\appendix

\section{Gaussian quadrature method}\label{sec:gqm}

The semi-analytic bounce calculation requires repeated evaluations of the
thermal functions $J_F$ and $J_B$. Standard high- and low-temperature
expansions are not uniformly accurate in the temperature range relevant for
the transition. We therefore use a two-point Gaussian quadrature rule for
the remaining radial integral. The thermal functions and their numerical
evaluation are discussed, for example, in \cite{Fowlie:2018eiu}, while the
general construction of Gaussian quadrature rules is reviewed in
\cite{GolubWelsch:1969}.

For $X=F,B$, we define
\begin{equation}
\Delta_X(T)
\equiv
\frac{m_X^2(0,T)}{T^2}.
\label{eq:Delta_X_def}
\end{equation}
In the notation used in the main text,
\begin{equation}
\Delta_F=\delta_f,
\qquad
\Delta_B(T)=\delta_b(T).
\label{eq:Delta_delta_relation}
\end{equation}
After inserting the Gaussian ansatz for the bounce profile and carrying out
the angular integration, the relevant thermal contribution can be written
as
\begin{equation}
F_X(\alpha,\Delta_X)
\equiv
2\pi
\int_0^\infty dx\,x^{1/2}
\left[
J_X\left(\Delta_X+\alpha e^{-2x}\right)
-
J_X(\Delta_X)
\right].
\label{eq:FX_radial}
\end{equation}
The normalization in \eq{eq:FX_radial} includes the factor $2\pi$ generated
by the radial integration. This convention is the one used in the main
text.

Introducing
\begin{equation}
w=e^{-2x},
\end{equation}
we obtain
\begin{equation}
F_X(\alpha,\Delta_X)
=
\frac{\pi}{\sqrt{2}}
\int_0^1 dw\,
\sqrt{-\ln w}\,
\frac{
J_X(\Delta_X+\alpha w)
-
J_X(\Delta_X)
}{w}.
\label{eq:FX_w_integral}
\end{equation}
Although the integrand in \eq{eq:FX_w_integral} contains an explicit factor
$1/w$, the apparent singularity at $w=0$ is removable:
\begin{equation}
\lim_{w\rightarrow0}
\frac{
J_X(\Delta_X+\alpha w)
-
J_X(\Delta_X)
}{w}
=
\alpha J_X^\prime(\Delta_X).
\label{eq:FX_endpoint}
\end{equation}

We now introduce the weight function
\begin{equation}
\rho(w)
=
\sqrt{-\ln w}
\label{eq:rho_weight}
\end{equation}
and approximate
\begin{equation}
\int_0^1 dw\,\rho(w)f(w)
\simeq
w_1 f(x_1)
+
w_2 f(x_2).
\label{eq:two_point_quadrature}
\end{equation}
The two nodes and weights are determined by requiring
\eq{eq:two_point_quadrature} to be exact for
$f(w)=1,w,w^2,w^3$. The corresponding moments are
\begin{equation}
\mu_n
\equiv
\int_0^1 dw\,w^n\sqrt{-\ln w}
=
\frac{\Gamma(3/2)}
{(n+1)^{3/2}}
=
\frac{\sqrt{\pi}}
{2(n+1)^{3/2}},
\qquad
n=0,1,2,3.
\label{eq:quadrature_moments}
\end{equation}
Hence,
\begin{align}
w_1+w_2
&=
\frac{\sqrt{\pi}}{2},
\\
w_1x_1+w_2x_2
&=
\frac{\sqrt{\pi}}{4\sqrt{2}},
\\
w_1x_1^2+w_2x_2^2
&=
\frac{\sqrt{\pi}}{2\,3^{3/2}},
\\
w_1x_1^3+w_2x_2^3
&=
\frac{\sqrt{\pi}}{16}.
\label{eq:quadrature_conditions}
\end{align}
Solving these equations gives
\begin{align}
x_1
&\simeq
0.1535900739,
&
x_2
&\simeq
0.6908657079,
\nonumber
\\
w_1
&\simeq
0.5563908709,
&
w_2
&\simeq
0.3298360546.
\label{eq:wixi}
\end{align}
Therefore,
\begin{equation}
F_X(\alpha,\Delta_X)
\simeq
\frac{\pi}{\sqrt{2}}
\sum_{i=1}^{2}
w_i
\frac{
J_X(\Delta_X+x_i\alpha)
-
J_X(\Delta_X)
}{x_i},
\qquad
X=F,B.
\label{eq:FX_quadrature}
\end{equation}

The derivative with respect to $\alpha$, which enters the stationary
conditions for the Gaussian bounce ansatz, is approximated by
\begin{equation}
F_X^\prime(\alpha,\Delta_X)
\simeq
\frac{\pi}{\sqrt{2}}
\sum_{i=1}^{2}
w_i
J_X^\prime(\Delta_X+x_i\alpha),
\qquad
X=F,B.
\label{eq:FXprime_quadrature}
\end{equation}
Here the prime on $F_X$ denotes differentiation with respect to $\alpha$,
while the prime on $J_X$ denotes differentiation with respect to its
argument.

The quadrature rule is exact for cubic polynomials with the weight
$\rho(w)=\sqrt{-\ln w}$. Since the thermal-function integrand is not a
polynomial, \eqs{eq:FX_quadrature}{eq:FXprime_quadrature} provide a
low-order approximation. Its numerical accuracy should be assessed by
comparison with direct integration over the parameter range used in the
bounce calculation.

\bibliographystyle{apsrev4-2}
\bibliography{phaseTransition}

\end{document}